\documentclass{article} 
\pdfoutput=1

\usepackage{jheppub}

\newcommand\fverb{\setbox\fverbbox=\hbox\bgroup\verb}
\newcommand\fverbdo{\egroup\medskip\noindent%
            \fbox{\unhbox\fverbbox}\ }
\newcommand\fverbit{\egroup\item[\fbox{\unhbox\fverbbox}]}
\newbox\fverbbox

\newcommand{\be}{\begin{eqnarray}}
\newcommand{\ee}{\end{eqnarray}}

\title{Searching for hidden mirror symmetries in CMB fluctuations
from WMAP 7 year maps}

\author[a,b]{Fabio Finelli}
\author[a,b]{Alessandro Gruppuso}
\author[c]{Francesco Paci}
\author[d]{and Alexey A. Starobinsky}

\affiliation[a]{INAF/IASF-BO, Istituto di Astrofisica Spaziale e Fisica Cosmica di Bologna,\\
    via Gobetti 101, I-40129 Bologna, Italy}
\affiliation[b]{INFN, Sezione di Bologna,\\ 
        Via Irnerio 46, I-40126 Bologna, Italy}
\affiliation[c]{Laboratoire Astroparticule \& Cosmologie (APC), CNRS, Universit\'e
    Paris Diderot, 10, rue A. Domon et L. Duquet,
    75205 Paris Cedex 13, France}
\affiliation[d]{Landau Institute for Theoretical Physics RAS, 119334 Moscow, Russia}

\emailAdd{finelli@iasfbo.inaf.it}
\emailAdd{gruppuso@iasfbo.inaf.it}
\emailAdd{fpaci@apc.univ-paris7.fr}
\emailAdd{alstar@landau.ac.ru}

\abstract{We search for hidden mirror symmetries
at large angular scales in the WMAP 7 year Internal Linear
Combination map of CMB temperature anisotropies using global pixel based estimators
introduced for this aim. Two different axes are found for which
the CMB intensity pattern is anomalously symmetric (or
anti-symmetric) under reflection with respect to orthogonal planes
at the $99.84(99.96)\%$ CL (confidence level), if compared to a result for an arbitrary axis in simulations without
the symmetry. We have verified that our results are robust to the introduction of
the galactic mask.
The direction of such axes is close to the CMB kinematic dipole and nearly orthogonal
to the ecliptic plane, respectively.
If instead the real data are compared to those in simulations taken with 
respect to planes for which the maximal mirror symmetry is generated
by chance, the confidence level decreases to $92.39 (76.65)\%$. But when the
effect in question translates into the anomalous alignment between
normals to planes of maximal mirror (anti)-symmetry and these natural axes
mentioned.
We also introduce the representation of the above estimators in the harmonic domain, confirming the 
results obtained in the pixel one.
The symmetry anomaly is shown to be almost entirely due to low
multipoles, so it may have a cosmological and even primordial
origin. Contrary, the anti-symmetry one is mainly due to
intermediate multipoles that probably suggests its non-fundamental
nature. We have demonstrated that these anomalies are not connected to the known issue of the 
low variance in WMAP observations
and we have checked that axially symmetric parts of these
anomalies are small, so that the axes are not the symmetry ones.}


\begin{document}

\maketitle


\section{Introduction}
\label{intro}

The simple and minimal $\Lambda$CDM standard cosmological model
with the approximately flat ($n_s\approx 1$) initial spectrum of
adiabatic density perturbations predicted by the inflationary
scenario of the early Universe is very successful in explaining
the observed Cosmic Microwave Background (CMB) properties of temperature and polarization anisotropies pattern
\cite{Komatsu:2010fb}, see also
\cite{Dunkley10} and \cite{Keissler11} for additional recent
high-$\ell$ data from the Atacama Cosmology Telescope and the South
Pole Telescope respectively, as mainly given
by statistically isotropic and Gaussian fields.
However, at the next level of
accuracy several anomalies have been found using very different
estimators. Though their contribution to the total {\em rms} values of
CMB temperature anisotropy and polarization is small, they seem
not to cope too easily with the statistical isotropy. At present it
is not clear if these anomalies have a common origin. Also their
statistical significance is under debate \cite{Bennett:2010jb}.

Namely, unlikely alignments of the quadrupole and the octupole were found in
\cite{Tegmark:2003ve,Copi:2003kt,Schwarz:2004gk,Weeks:2004cz,Land:2005ad} for the WMAP first year release maps.
Both quadrupole and octupole were shown to align with the CMB dipole \cite{Copi:2005ff,Copi:2006tu} (other unlikely
alignments are described in \cite{Abramo:2006gw,Wiaux:2006zh,2007MNRAS.381..932V,Gruppuso:2010up}
and a test for detecting foreground residuals in the alignment estimators is proposed in \cite{Gruppuso:2009ee}).
Also, it is found that the power in the CMB temperature anisotropy coming separately
from the two hemispheres (defined by the ecliptic plane) is unlikely symmetric \cite{Eriksen:2003db,Hansen:2004vq}.
It has been confirmed in the WMAP 3 year and 5 year release \cite{Eriksen:2007pc,Hansen:2008ym,Hoftuft:2009rq}
and it is present in the COBE data as well, although with lower significance.
The temperature power spectra of the opposing hemispheres are inconsistent at $3\sigma$ to $4\sigma$ depending on the
range of multipoles considered. The asymmetry has been detected in low resolution maps \cite{Eriksen:2003db}, both in angular and multipoles
space, but it extends to much smaller angular scales in the multipole
range $\delta \ell = [2,600]$ \cite{Hansen:2008ym}.
At large angular scales the hemispherical asymmetry has been tested for the first time in polarization maps in \cite{Paci:2010wp}
making clear that this anomaly is mainly evident only in intensity at WMAP sensitivity.

It has been suggested in \cite{Land:2005jq} that an estimator
built upon the point parity symmetry might be used as a practical tool
for detecting foregrounds.
In particular the authors consider whether the observed low
CMB quadrupole in temperature could more generally signal odd point-parity, i.e. suppression of even multipoles.
However, they claim that WMAP 1st year data does not support parity preference beyond the meagre $95 \%$ confidence level.
Later, it was found \cite{Kim:2010gf} that the parity symmetry in the temperature map of WMAP 3 and 5 year data is
anomalous at the level of 4 out of 1000 in the range $\delta \ell = [2,18]$.
Other analysis in the WMAP 7 year data confirm the anomaly at same level for a slightly wider range
$\delta \ell = [2,22]$ \cite{Kim:2010gd} and show that the anomaly has a characteristic scale around
$\ell \sim 20$ \cite{Gruppuso:2010nd} .
See also \cite{Kim:2010zn,Hansen:2011wk} for extensions of such analysis.

Existence of these anomalies shows that CMB temperature fluctuations (as well as polarization ones) may contain
some additional hidden information (pattern, "inscription") not evident from the standard analysis because it is
masked by the dominant, statistically isotropic and Gaussian component. If related to low multipoles, this
information can provide us with some completely unexpected knowledge about early stages of the evolution of our
Universe including its very origin. However, it is not easy to extract it without having at least some crude idea
of how it may look like. A hidden partial symmetry of CMB fluctuations, i.e. the existence of a subcomponent of CMB
fluctuations for which this symmetry is an exact one, is one of the simplest variants of such hidden information.

In particular, this symmetry can be the mirror one with respect to
reflection in some plane. As was shown in Ref. \cite{Starobinsky},
one physical mechanism leading to the appearance of a subcomponent
of CMB angular temperature fluctuations possessing such symmetry
is the existence of a non-trivial spatial topology: the
non-compact $T^1$ topology, or the compact $T^3$ topology in which
one of its topological scales is much less than two others (also
called a slab space). For such spatial topology, the corresponding
component of wave vectors of perturbations, say $k_z$, is
quantized, and then the subcomponent with the exact mirror
symmetry with respect to reflection in the XY plane is produced by
perturbations with $k_z=0$. However, it is evident that a similar
effect can arise even in the absence of any non-trivial spatial
topology but if, for some (may be pure statistical) reason,
large-scale scalar (density) metric perturbations in the region of
space including our last scattering surface have a significant
component with a very weak dependence on one of the spatial
coordinates.

This type of partial symmetry was first searched for in Ref.
\cite{deoliveirasmootstaro} using the COBE data with the negative
result placing a lower limit of the minimal present comoving
size $L$ of the torus of the order of $4$ Gps. After that this
limit has been raised much, and now is close to $2R_{ls}\approx
28$ Gps for the cubic $T^3$ topology obtained using the most
recent WMAP data and searching for identical circles in the sky
\cite{Bielewicz:2010bh} ($R_{ls}$ is the present comoving radius
of our last scattering surface). Note, however, that a partial
mirror symmetry can be seen in CMB fluctuations even for
$L>2R_{ls}$ when there is no identical circles in CMB
fluctuations.

For this reason, in the present paper we have performed a new search for a hidden mirror reflection symmetry in the
WMAP 7 year temperature maps developing estimators in the pixel domain and in the harmonic domain (the latter only
for the full sky case). For generality, we look for a mirror reflection anti-symmetry, too, though it is less clear
how such symmetry may arise. The paper is organized as follows. In Section \ref{estimators} we define the used tools
to perform the analysis reported in Section \ref{analysis}. We discuss our results in connection with the issue 
of the low variance in WMAP observations in Section \ref{discussion}.
Our conclusions are drawn in Section \ref{conclusions}.
Algebraic details for the equivalence between the estimators in harmonic and pixel domains are given in Appendix
\ref{appendix}.

\section{Tools and estimators}
\label{estimators}

We consider the following pixel based estimators
\be
Q(\hat n_i) =  - {1 \over N_{\rm pix}} \sum_{j=1}^{N_{\rm pix}}
{\delta T \over T} (\hat n_j) {\delta T \over T}(\hat n_k)  = S^{-} (\hat n_i) - S^{+}(\hat n_i)
\label{defQ}
\, ,
\ee
where $S^{\mp} (\hat n_i)$ are defined in analogy to \cite{deoliveirasmootstaro} as follows
\be
S^{\mp} (\hat n_i) = {1 \over N_{\rm pix}} \sum_{j=1}^{N_{\rm pix}}
\left[ {1 \over 2} \left( {\delta T \over T} (\hat n_j) \mp {\delta T \over T}(\hat n_k) \right)\right]^2
\label{defSpem}
\, ,
\ee
where the sum is meant over the observed pixels, $N_{\rm pix}$, ${\delta T / T} (\hat n_j)$ are the temperature CMB anisotropies measured
at the pixel pointed by the unit vector $\hat n_j$ and $\hat n_k$ is the opposite direction of $\hat n_j$ with respect
to a plane defined by $\hat n_i$,
i.e.
\be
\hat n_k = \hat n_j -2\, ( \hat n_i \cdot \hat n_j) \hat n_i
\, .
\ee
We stress that only two out of the three global estimators given in Eqs.~(\ref{defQ}),(\ref{defSpem}) are independent
and that $S^{\pm}$ is more sensitive to increased angular resolution than $Q$ (see below).
Eqs.(\ref{defQ}) and (\ref{defSpem}) can be rewritten in the harmonic space as follows (see Appendix \ref{appendix} for details):
\be
Q(\hat n_i) &=& - {1 \over 4 \pi} \sum_{\ell} \left[ (\ell+1) \, C^{(+)}_{\ell} -  \ell \, C^{(-)}_{\ell} \right] \, \label{defQh} , \\
S^{\pm} (\hat n_i) &=& {1 \over 2} {1 \over 4 \pi} \sum_{\ell}
\left\{ (2 \ell +1) C_{\ell} \pm \left[  (\ell+1) \,
C^{(+)}_{\ell} -  \ell \, C^{(-)}_{\ell} \right] \right\} \, ,
\label{defSpemh} \ee where $(\ell+1) \, C^{+}_{\ell} = \sum_{m=-\ell}^\ell
p^{+}_{\ell m} |a_{\ell m}|^2$ and $\ell \, C^{-}_{\ell} =
\sum_{m=-\ell}^\ell p^{-}_{\ell m} |a_{\ell m}|^2$ with $p^{\pm}_{\ell m}
= 1$ if $(\ell + m)$ is even/odd and $p^{\pm}_{\ell m} = 0$ if
$(\ell + m)$ is odd/even. The estimators $C^{\pm}_{\ell}$ depend
on the direction $\hat n_i$ since the coefficients of spherical
harmonics $a_{\ell m}$ are obtained in the frame in which the
$z-$axis lies along the direction defined by $\hat n_i$. Notice
that the quantity $S^+ + S^-$ is always direction independent if
computed over the whole sky, and it equals the total dispersion of
the CMB fluctuations: \be S^+ + S^- = {1 \over N_{\rm pix}}
\sum_{N_{\rm pix}} \left( \frac{\delta T} {T} \right)^2 =
\frac{1}{4\pi}\sum_{\ell}(2l+1)C_{\ell} \ee where $C_{\ell}$ is
the angular power spectrum of CMB anisotropies (see Appendix
\ref{appendix}).\footnote{Note that our estimator $Q$ differs
significantly from the one adopted in the recent paper
\cite{BenDavid:2011fc}
which does not have a simple expression in the pixel space.}

If different CMB anisotropy multipoles $a_{lm}$ are not correlated and statistically isotropic,
then the average values of $Q(\hat n_i)$, $S^+(\hat n_i)$ and $S^-(\hat n_i)$ are given by
\be
\langle Q(\hat n_i) \rangle &=& - {1 \over 4 \pi} \sum_{\ell} C_{\ell}  \label{Qmean} \\
 \langle S^-(\hat n_i) \rangle &=& {1 \over 4 \pi} \sum_{\ell} \ell \, C_{\ell} \label{Spmean}\\
 \langle S^+(\hat n_i) \rangle &=& {1 \over 4 \pi} \sum_{\ell} (\ell +1) \, C_{\ell} \label{Smmean}
\ee
The r.h.s. of Eqs.~(\ref{Qmean}), (\ref{Spmean}) and (\ref{Smmean}) are invariant under rotation, i.e.
independent of the direction $\hat n_i$, as it should be.
On the other hand, if CMB fluctuations contain a significant sub-component exactly symmetric (anti-symmetric)
with respect to reflection in a plane with the normal vector $n_i$, the estimator $S^-(n_i)~(S^+(n_i))$ will be
anomalously small while the estimator $S^+(n_i)~(S^-(n_i))$ anomalously large. The maximum and minimum values
for $S^-$ are also maximum and minimum values for $Q$, respectively.

\section{The Analysis}
\label{analysis}

\subsection{Data, Simulations and Results for the pixel based estimators}
\label{pixelstatistics}

We have considered the WMAP 7 yr Internal Linear Combination (ILC) map at
Healpix\footnote{http://healpix.jpl.nasa.gov} resolution $N_{\rm side}=16$ (corresponding to maps composed by $3072$ pixels,
\cite{gorski}), smoothed at angular resolution FWHM $= 9.1285^\circ$. For each direction defined by the center of each
pixel at such resolution, we have computed the estimators $S^+$, $S^-$
and $Q$ defined by Eqs.~(\ref{defQ}) and (\ref{defSpem}).
We have then repeated the same procedure for $10000$ full sky CMB maps,
extracted from the best fit of WMAP 7 yr $\Lambda$CDM model
\cite{Komatsu:2010fb} at the same resolution and smoothed at the
same angular scale as the WMAP map.
Since random extractions are independent, CMB anisotropies are isotropically distributed on the sphere and the distributions
of the estimators given in Eqs.~(\ref{defQ}) and (\ref{defSpem}) do not depend on the direction $\hat n_i$ for the full sky case.

\begin{figure}
\includegraphics[width=4.8cm]{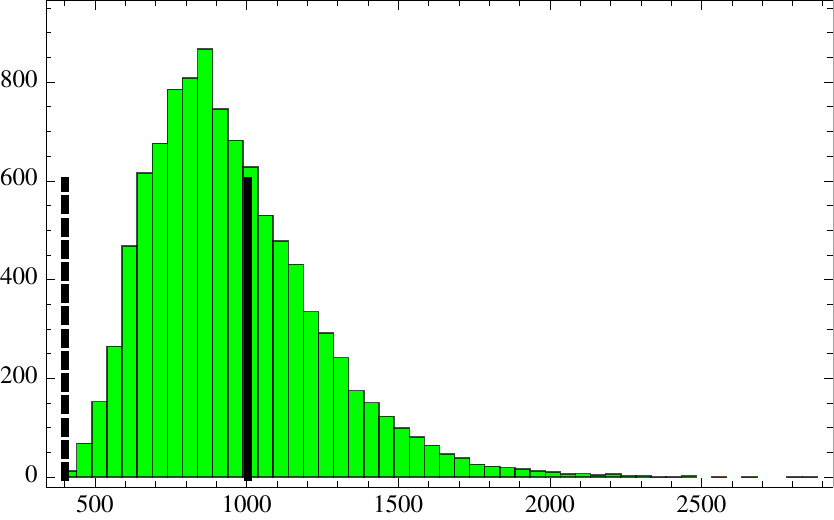}
\includegraphics[width=4.8cm]{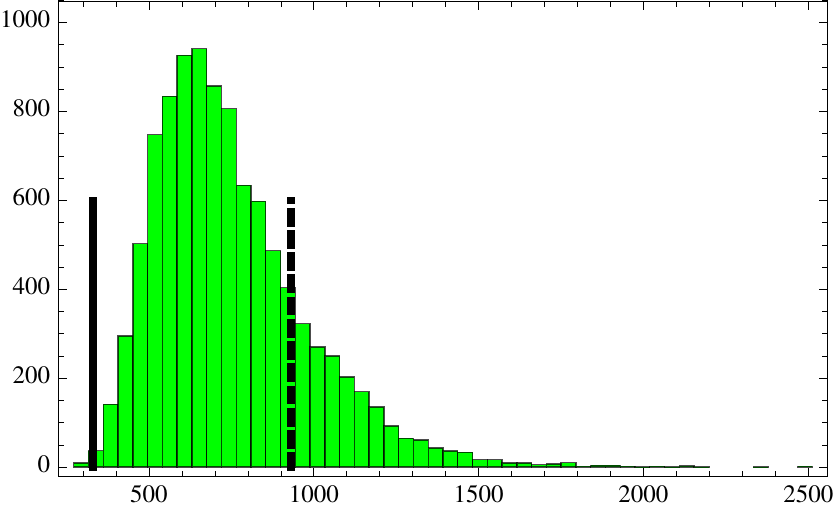}
\includegraphics[width=4.8cm]{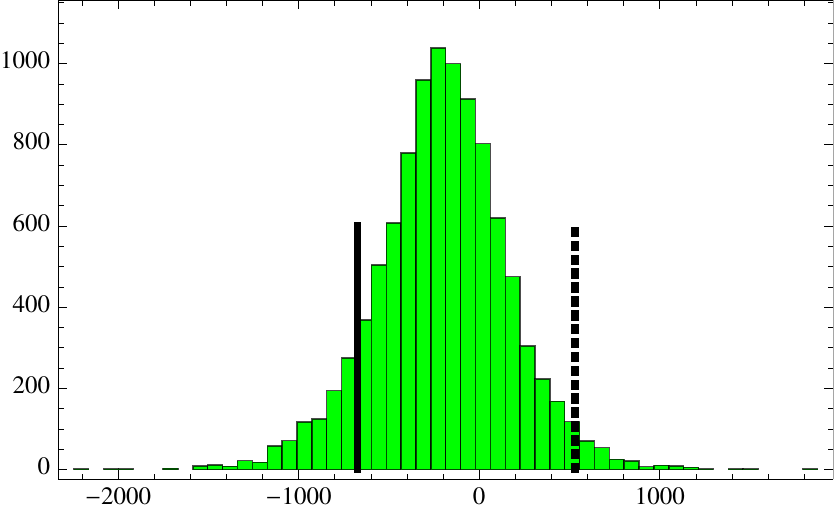}
\caption{Distribution of $S^{+}$ (left panel), $S^{-}$ (middle panel)
  and $Q$ (right panel) for the full sky analysis. For each estimator,
  vertical lines show
  the most anomalous values on the
  WMAP 7 yr ILC map: solid lines correspond to $\hat{d}_2$, whereas dashed
  lines correspond to $\hat{d}_1$. The green histograms are obtained using $10000$ Monte Carlo realizations for $S^+ (\hat{d}_1)$ (left panel) 
and $S^- (\hat{d}_3)$ (right panel).}
\label{Pixelbased}
\end{figure}

Among all the directions connected to the HEALPIX pixelization at $N_{\rm side}=16$, for WMAP 7 ILC map we find that the
maximum value for $S^+$ corresponds to the minimum value for $S^-$, and vice versa. The minimum and maximum value of
$S^+$ are plotted in Fig.~\ref{Pixelbased}
(dashed and solid vertical lines) on top of the distributions for the same quantities resulting from our $10000$
Monte Carlo realizations (green histograms).
The minimum values for $S^\pm$ obtained by the WMAP 7 yr full sky ILC map are quite small compared to the probability distribution obtained by
the Monte Carlo method, whereas their maximum values are not anomalous. The maximum and minimum values for $Q$ do not exhibit anomalies 
with respect
to the Monte Carlo distribution.
We have explicitly checked that the probability distributions of
$S^\pm$ computed on WMAP 7 yr full sky ILC map over the 3072 directions associated to the HEALPIX $N_{\rm side}=16$ differ
from the corresponding ones obtained from a random Gaussian CMB realization. The anomalous minima values for $S^\pm$ are a consequence
of the corresponding anomalous distributions obtained on WMAP 7 yr ILC map.

It turns out that $S^+$ reaches its minimum value for the direction
$\hat{d}_1$ defined by $(\theta=107^\circ,\phi=264^\circ)$, direction for which
$S^-$ is maximum.
Instead, $S^-$ is minimized along
$\hat{d}_2$, defined by $(\theta=42^\circ,\phi=260^\circ)$, which also corresponds
to the maximum of $S^+$. Throughout the present paper, spherical coordinates
are given in ($\theta,\phi$), being $\theta$ and $\phi$ the galactic co-latitude and
longitude respectively\footnote{Spherical
  coordinates are sometimes provided in the literature in terms of
  galactic longitude and latitude, $(l,b)$, related to
  the coordinates adopted in this paper by  $l=\phi$ and $b=90^\circ-\theta$. The two
  directions $\hat{d}_1$ and $\hat{d}_2$ would therefore read
  $(l=264^\circ,b=-17^\circ)$ and $(l=260^\circ,b=48^\circ)$ respectively.}.
For comparison, the direction corresponding to the hemispherical power asymmetry is
$(\theta=107^\circ,\phi=226^\circ)$ \cite{Hansen:2008ym}, while that of the
kinematic dipole measured by WMAP is $(\theta=42^\circ,\phi=264^\circ)$ \cite{Jarosik:2010iu}
that is very close to the direction $\hat{d}_2$.


The probabilities related to the directions considered are summarized
in Table \ref{tableprobabilities}. The most significant departure
from symmetry is the $S^+$, minimum along the direction
$\hat{d}_1$, followed by the $S^-$, minimum along
the direction $\hat{d}_2$. The estimator $Q$ shows no significant anomalies with respect to our
Monte Carlo realizations. The secondary minimum in $S^+$ - $S^-$ and $Q$ - map is placed in the
direction characterized by $(\theta = 90^\circ \,, \phi = 200^\circ)$  - $(\theta =  171^\circ \,, \phi = 165^\circ)$ -.
The probabilities related to these second minima are $99.70 \% \,, 99.65
\% \,, 89.42 \%$ for $S^+ \,, S^- \,, Q$ maps, respectively.

\begin{table}[ht]
\caption{Probabilities (in percentage) to obtain a larger value for the considered directions and pixel based estimators. Numbers in brackets refer to directions obtained using the masked analysis.} 
\centering 
\begin{tabular}{c c c c c} 
\hline\hline 
Estimator / Direction & $\hat{d}_1$ &  $\hat{d}_2 (\hat{d}_3)$\\ [0.5ex] 
\hline 
$S^{+}$ & 99.99 ($>$99.99) & 36.58\\ 
$S^{-}$  & 18.78 & 99.86 (99.99)\\
$Q$ & 2.35 & 90.67 \\  [1ex] 
\hline 
\end{tabular}
\label{tableprobabilities} 
\end{table}

\begin{figure}
\includegraphics[width=7.0cm]{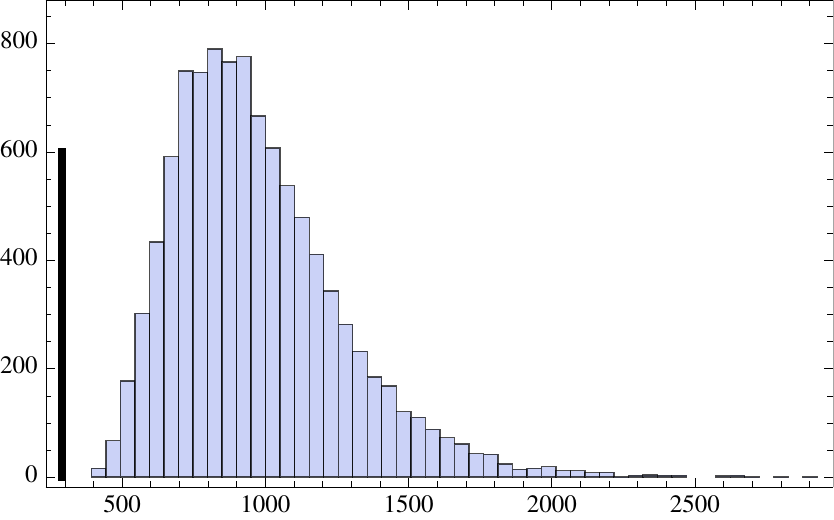}
\includegraphics[width=7.0cm]{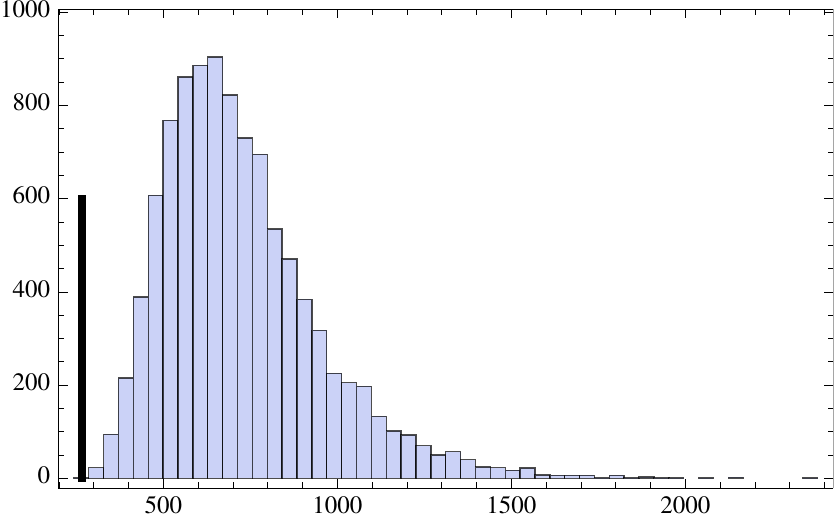}
\caption{Distribution of $S^+$ (left panel) and $S^-$ (right panel) for
the masked sky analysis. $S^+$ is minimized along the same direction
found over the full sky, $\hat{d}_1$. The minimum of $S^-$ points
instead along direction $\hat{d}_3$, with a small shift compared to
the full sky result, $\hat{d}_2$. The blue histograms are obtained 
using $10000$ Monte Carlo realizations for $S^+ (\hat{d}_1)$ (left panel) 
and $S^- (\hat{d}_3)$ (right panel) with the galactic mask.}
\label{Pixelbasedmasked}
\end{figure}

\begin{figure}
\includegraphics[angle=180,width=14.0cm]{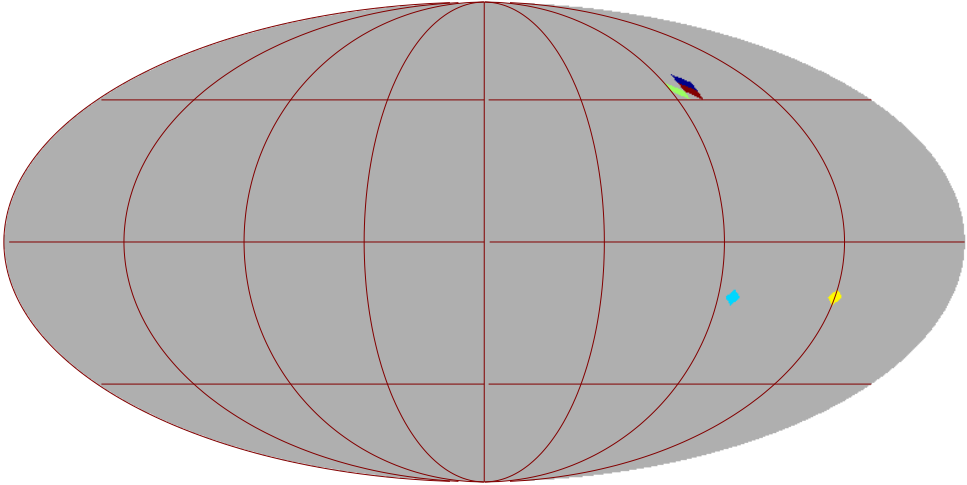}
\caption{Sky projection of the relevant directions: light blue for $\hat{d}_1$; red for
  $\hat{d}_2$; blue for $\hat{d}_3$; yellow for the N/S asymmetry;
  green for the kinematic dipole}
\label{directions}
\end{figure}

We have also performed the same analysis by masking the galactic plane and considering in the sums
only the contribution from those pixels which are not masked together with its opposite. The factor $1/N_{\rm pix}$
in Eqs. (2.1,2.2) takes into account only these pixels.
The comparison of the anomalous directions between full sky and masked sky shows that they
are stable
at the chosen resolution (see Fig.~\ref{directions}), as only a minor
deviation appears for the minimum of $S^-$ along the direction
$\hat{d}_3$ of coordinates $(\theta=39^\circ,\phi=260^\circ)$. However, the level of anomaly
increases for the masked sky for both $S^+$ and $S^-$ 
(see Table \ref{tableprobabilities} and figure \ref{Pixelbasedmasked}).

\begin{figure}
\includegraphics[width=4.8cm]{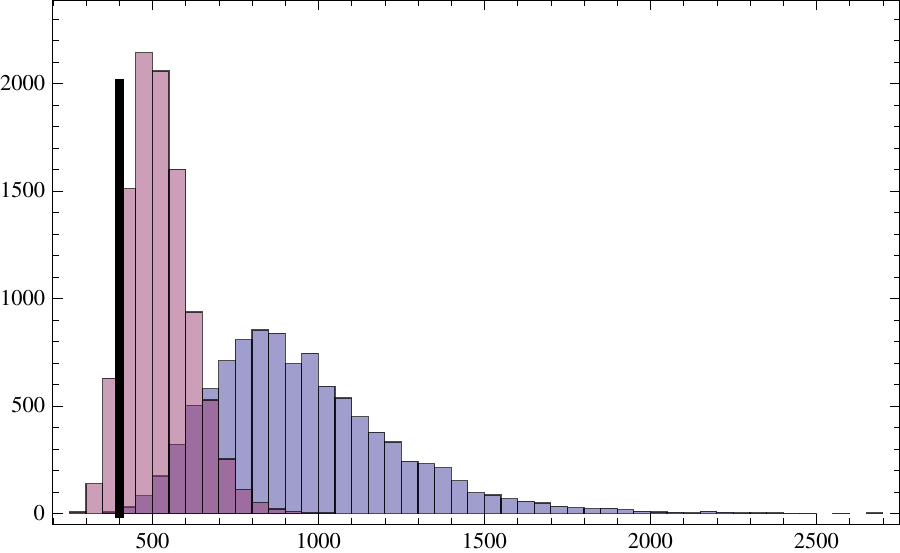}
\includegraphics[width=4.8cm]{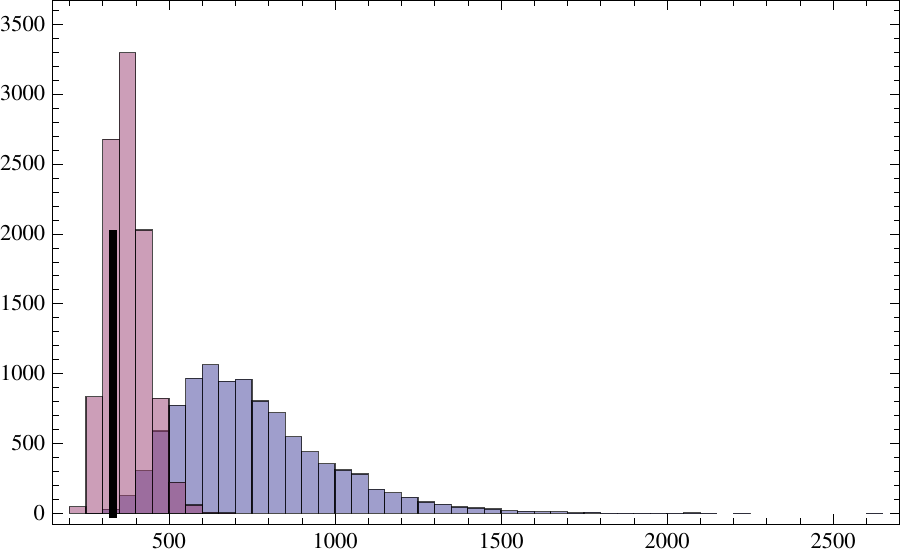}
\includegraphics[width=4.8cm]{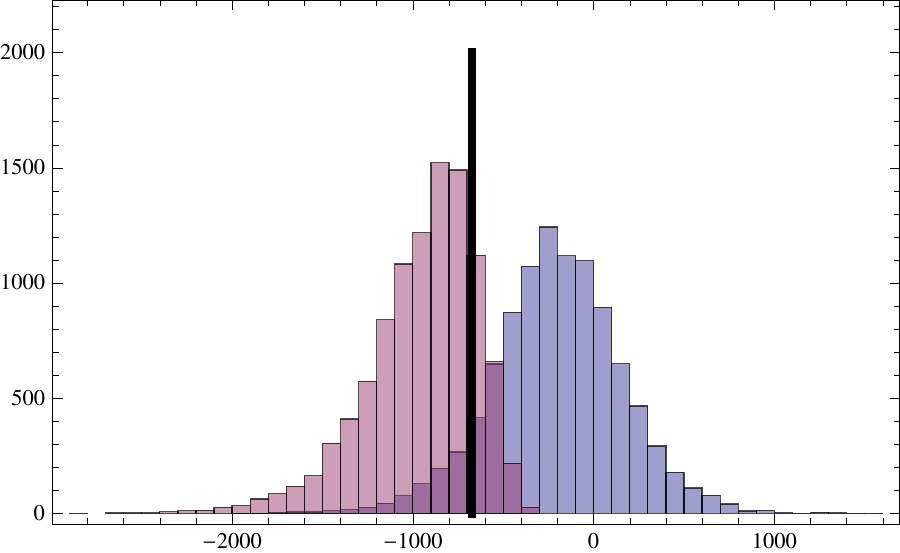}
\caption{Distribution of $S^+$ (left panel), $S^-$ (middle panel) and $Q$ (right panel) for the full sky analysis.
The most anomalous values on the WMAP ILC map (vertical line) is compared to two sets of MC simulations:
in violet the estimators for a fixed axis (as in Fig. \ref{Pixelbased}) and in pink the minimum value across 
the sky computed independently for each sky.}
\label{absolute}
\end{figure}

So far we have reported probabilities based on a comparison between real data and simulations without symmetries 
with respect to a fixed axis. If instead the real data are compared to the minimum of $S^\pm$ across the sky in each of 10000 random 
simulations, the statistical significance drops to $92.39 (76.65) \%$ for $S^+$ ($S^-$) on the WMAP 7 yr ILC map, as can be seen in 
Fig. \ref{absolute}; $14.77 \%$ of the simulations 
lie to the left of the the minimum value for $Q$ on the WMAP 7 yr ILC. On the basis of this second comparison of absolute minimum values of estimators on 
real data vs on simulations across the sky for each random realization, 
the anomaly translates into the question of an anomalous alignment between the normals 
to the planes of maximal mirror symmetry of real data with known directions such as the normal to the ecliptic plane and the cosmological dipole.

\begin{figure}
\includegraphics[width=6.8cm]{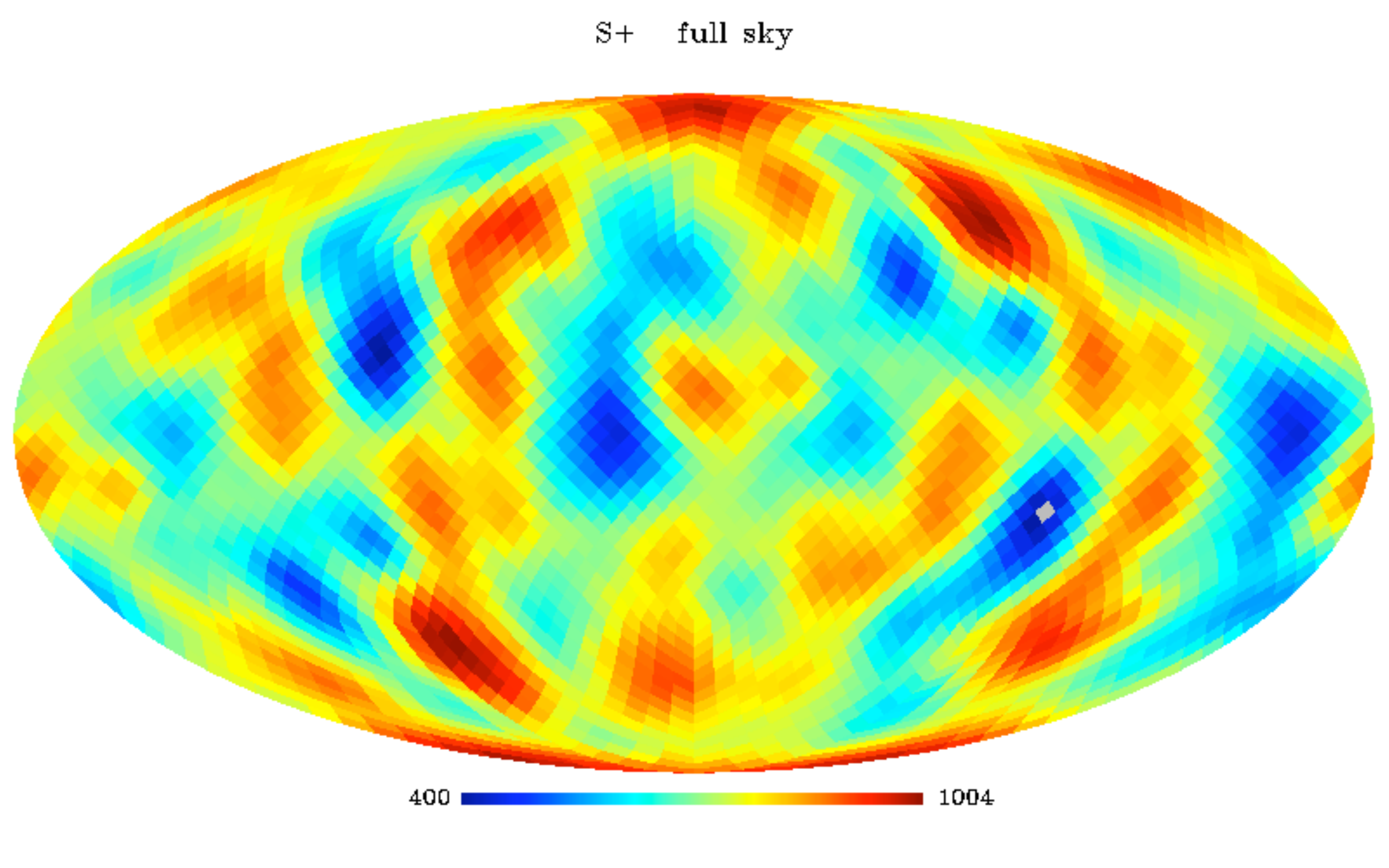}
\includegraphics[width=6.8cm]{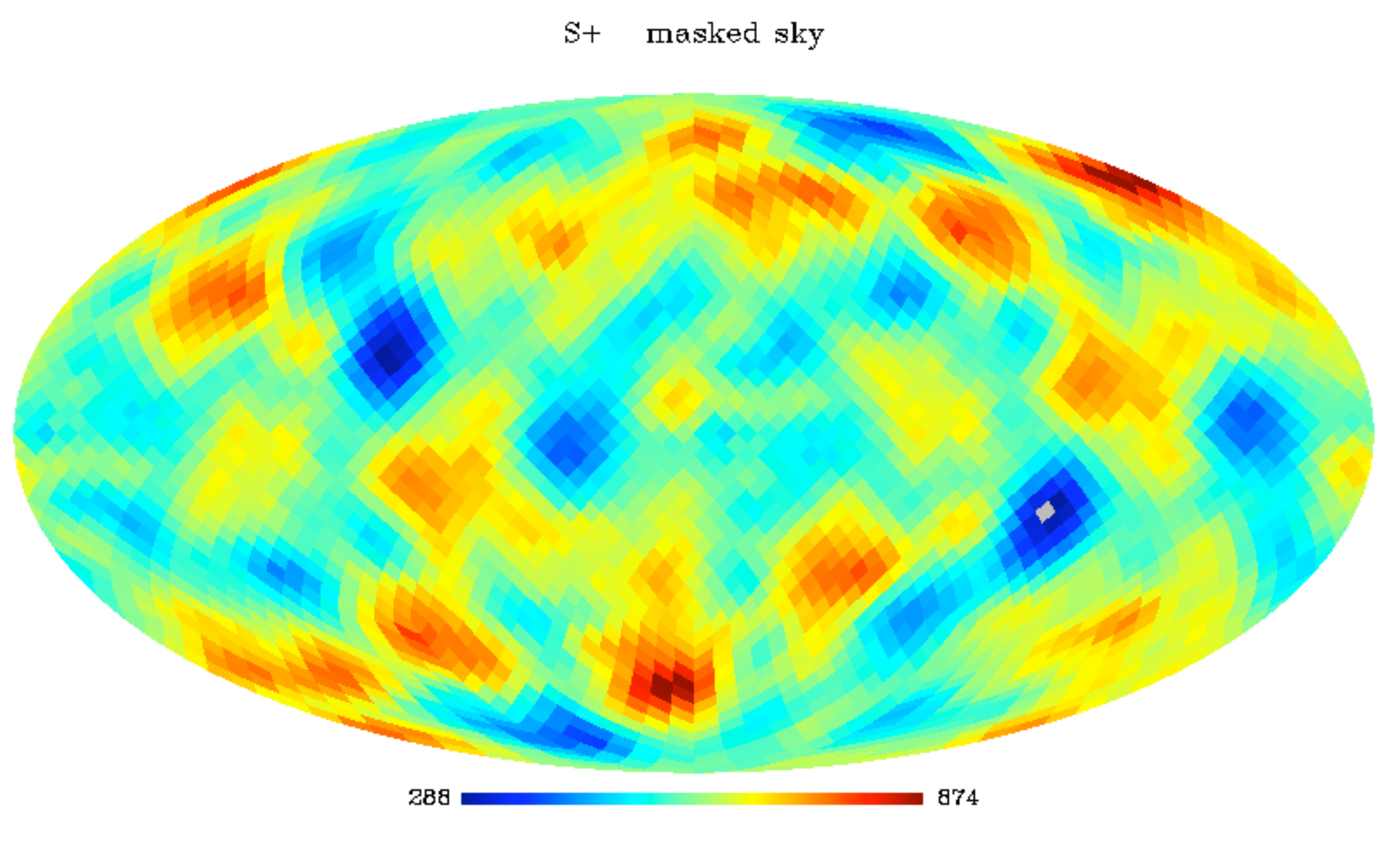} \\
\includegraphics[width=6.8cm]{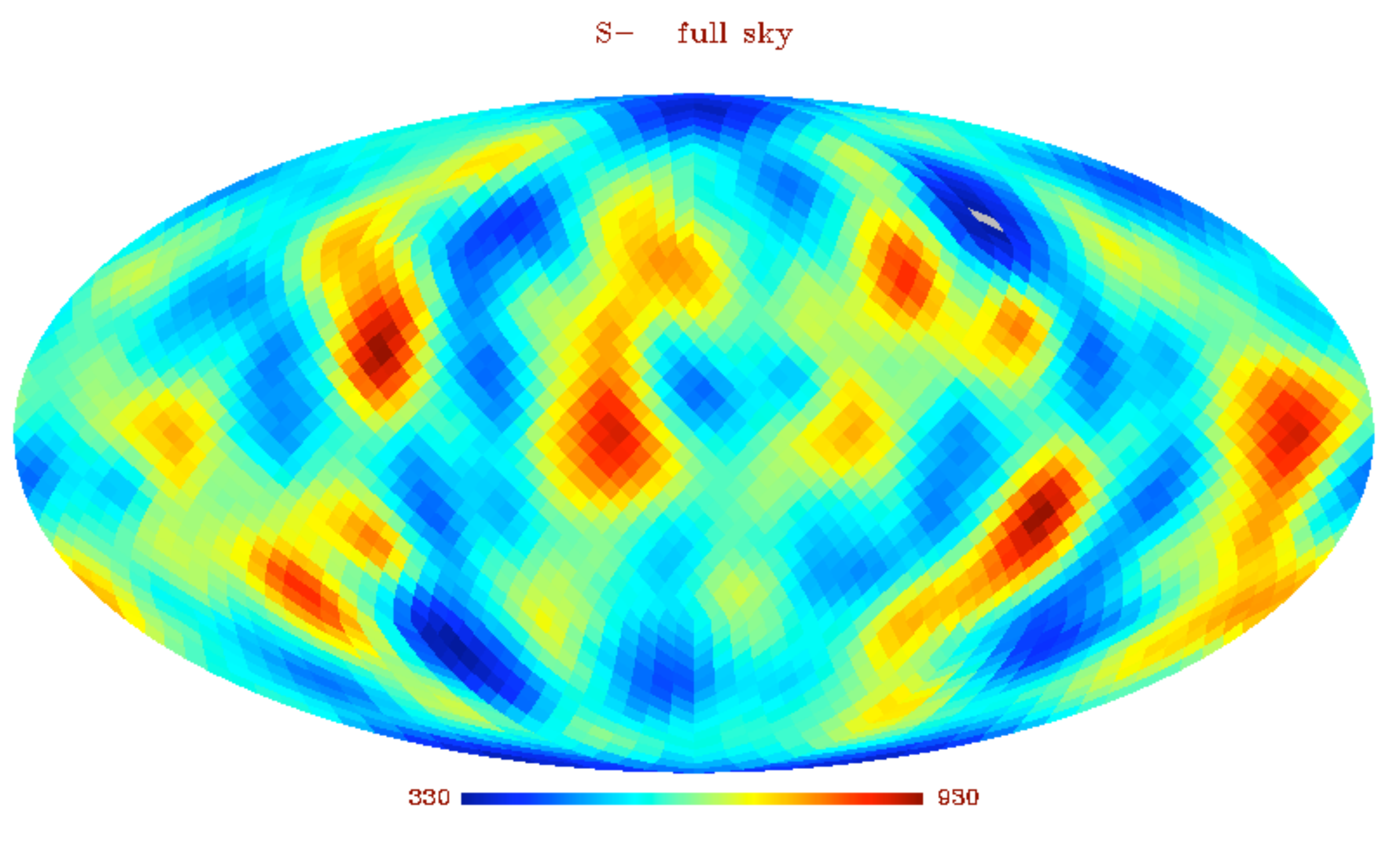}
\includegraphics[width=6.8cm]{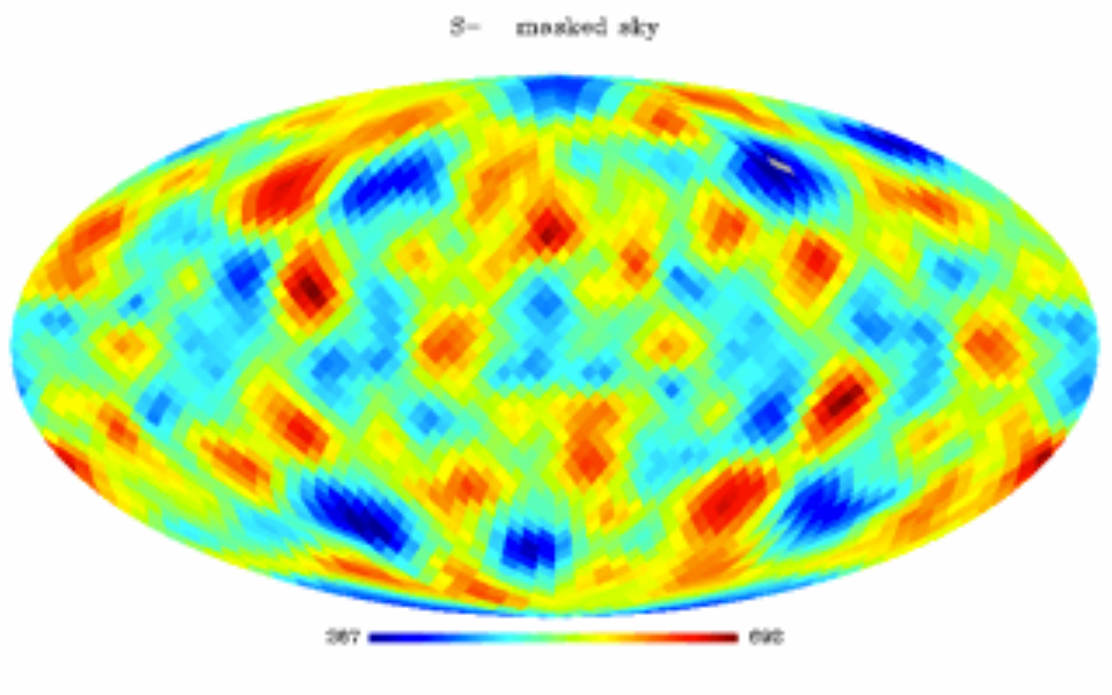} \\
\includegraphics[width=6.8cm]{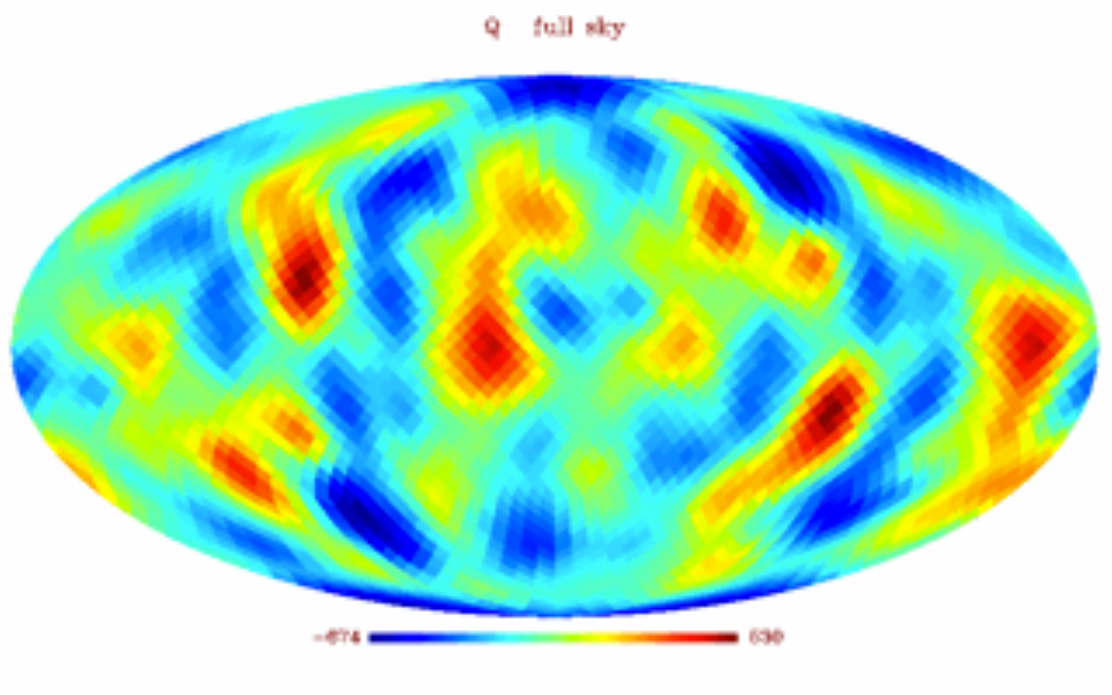}
\includegraphics[width=6.8cm]{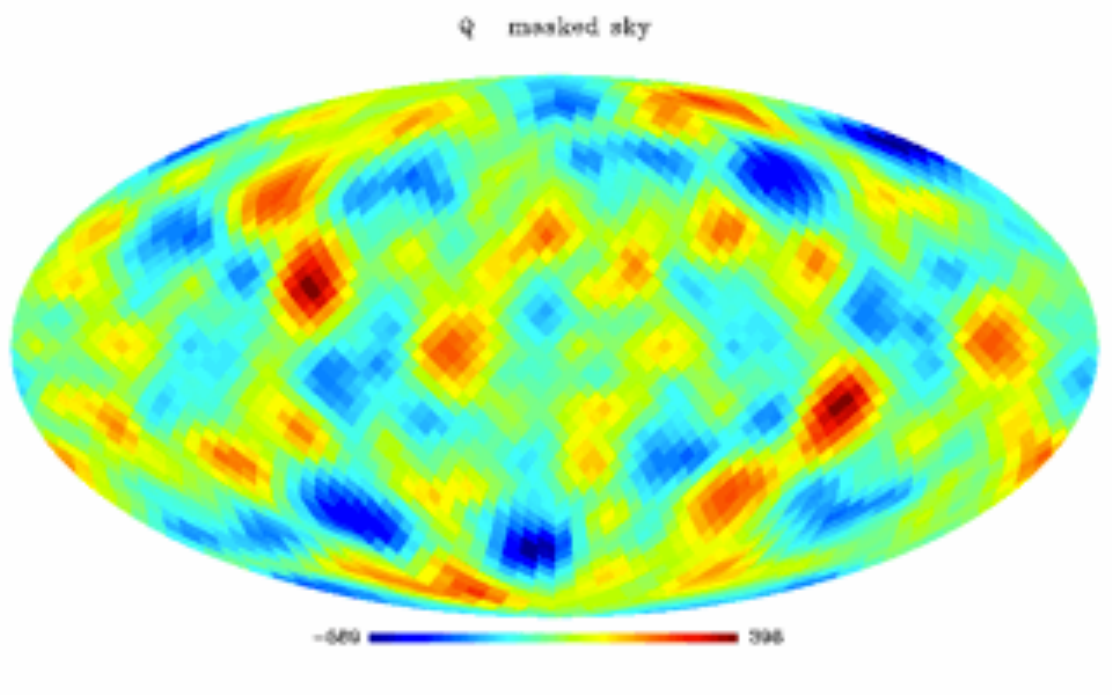}
\caption{Map of $S^+$ (top), $S^-$ (middle) and $Q$ (bottom) for the full (left) and masked (right) case.}
\label{score}
\end{figure}

In Fig. \ref{score} 
we provide now the full sky score map for $Q$, i.e. $Q (\hat n_i)$ in Eq. (\ref{defQ}).
Very similar maps are the full sky score maps 
for $S^\pm$, i.e. $S^\pm (\hat n_i)$ in Eq. (\ref{defSpem}) (where the two anomalous 
directions are shown in grey), which are given by
$\pm Q$ shifted by the value of the variance, which is represented by a 
monopole in these maps since it is direction independent. The full sky maps are 
even under (point) parity symmetry.
The score maps are shown in Fig. \ref{score} also for the 
masked case, although the masked score maps for $S^\pm$ and $Q$ are 
all different.


\subsection{Data, Simulations and Results for the harmonic based estimators}
\label{harmonicstatistics}

In order to study the angular scale dependence of the two anomalous directions found through the application
of the pixel based estimators, we have considered the harmonic version of the same estimators, given
in Eqs.~(\ref{defQh}) and (\ref{defSpemh}).

\begin{figure}
\includegraphics[width=7.5cm]{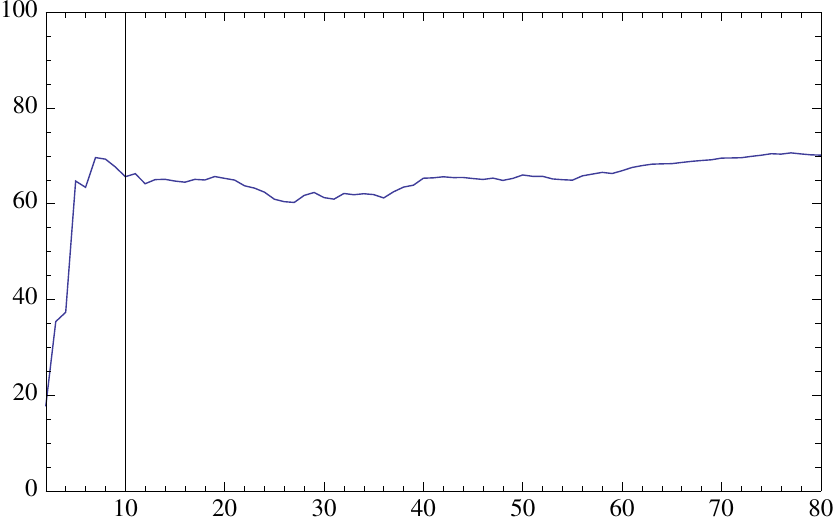}
\includegraphics[width=7.5cm]{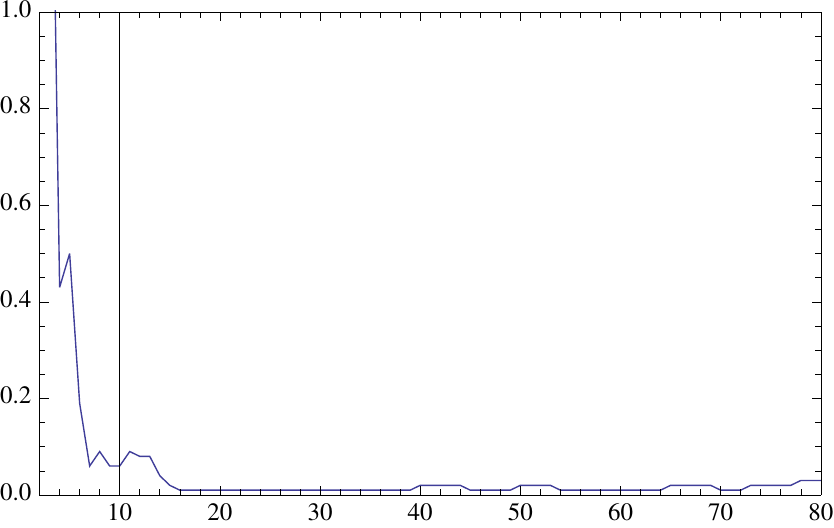}
\includegraphics[width=7.5cm]{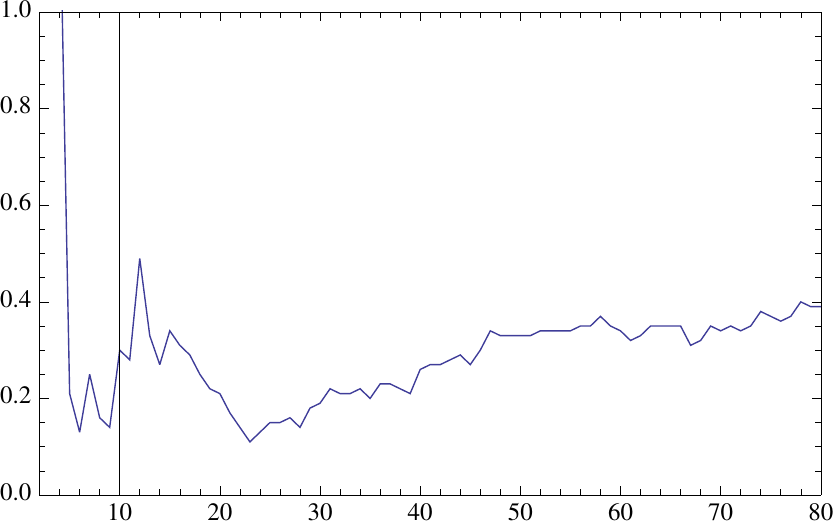}
\includegraphics[width=7.5cm]{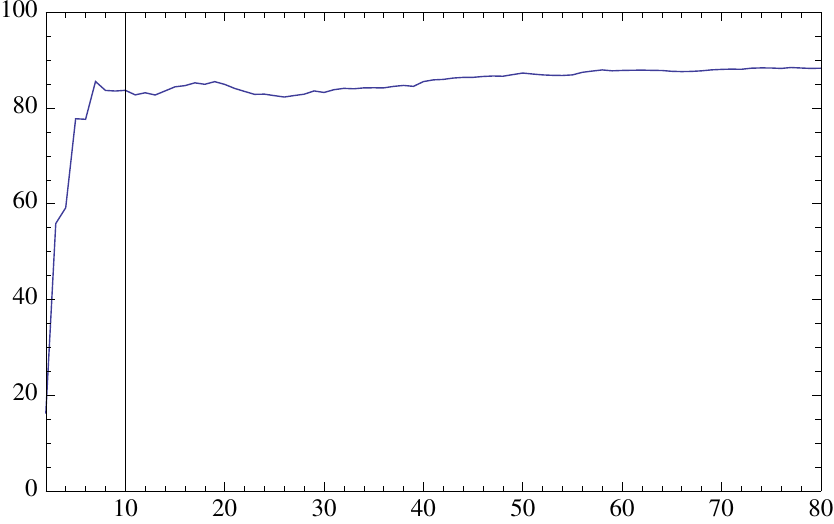}
\includegraphics[width=7.5cm]{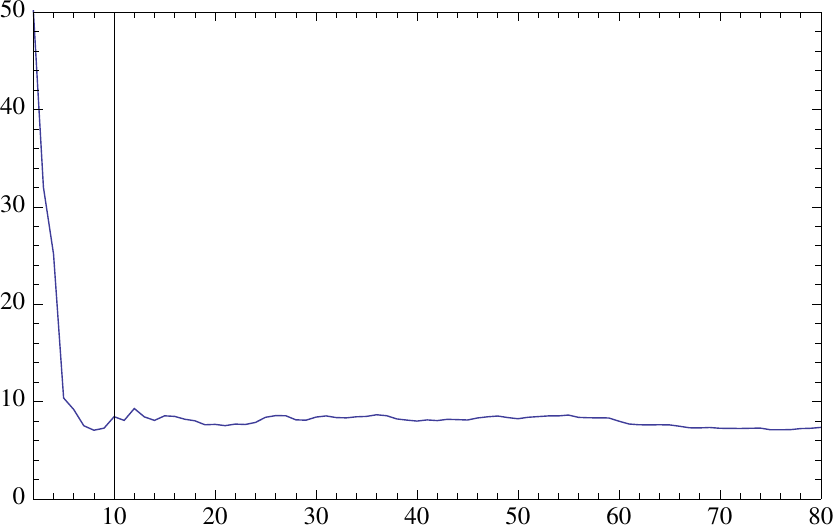}
\includegraphics[width=7.5cm]{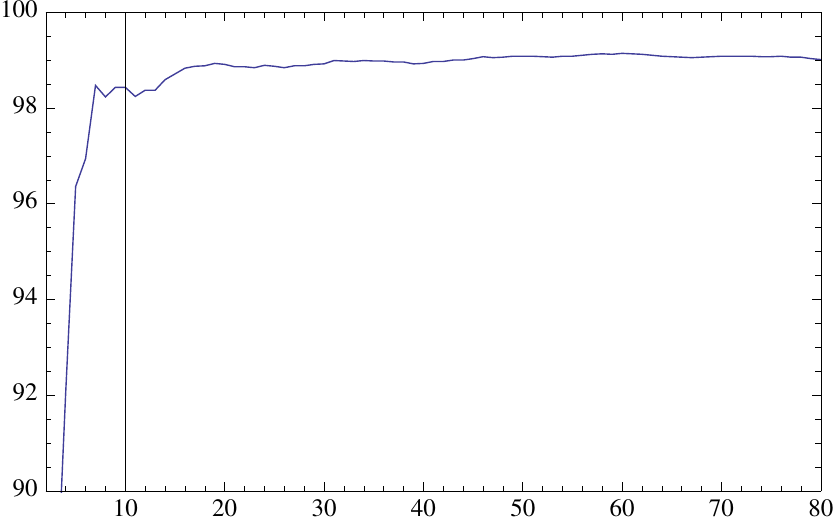}
\caption{Percentage versus the multipole number $\ell_{\rm max}$. For each $\ell_{\rm max}$ and for each estimator, we plot the
percentage to obtain a larger value w.r.t. the WMAP ILC one.
Left column is for the $\hat{d}_2~(42^\circ,260^\circ)$ direction and right column is for the $\hat{d}_1~
(107^\circ,264^\circ)$ direction.
First row for $S^{+}$, second row for $S^{-}$ and third row for $Q$.}
\label{percentage}
\end{figure}

We have smoothed the WMAP 7 yr full resolution ILC map at angular
resolution FWHM $= 2.2903^\circ$ and reconstructed at lower resolution $N_{\rm
  side}=64$. We have then extracted 10000 CMB realizations at the same
resolution $N_{\rm side}=64$ from the best fit of
the WMAP 7 year $\Lambda$CDM model and smoothed at the same angular
scale as the WMAP map.
For each of the two directions found in the pixel-based analysis, we
have computed $S^+$, $S^-$ and $Q$ as in Eqs.~(\ref{defQh}) and
(\ref{defSpemh}), for both WMAP and simulations, substituting the indefinite sum over a multipoles with a
definite one between $\ell_{\rm min}=2$ to $\ell_{\rm max}=80$.
By comparing those, we have computed the probability of WMAP with
respect to the random and independent CMB Gaussian extractions.
Results are plotted in Fig.~\ref{percentage}. The most anomalous case turns out to be the $S^{+}$
estimator for the direction $\hat{d}_1$
and for $\ell_{\rm max} > 15$ where the probability of getting such a low value is $0.01\%$, see upper right panel
of Fig.~\ref{percentage}. The estimator $S^{-}$ for the direction $\hat{d}_2$
is, on
average over $l$, anomalous at the level of $\sim 0.3\%$ for $\ell_{\rm max} > 4$
with a peak reaching $0.13 \%$ for $\ell_{\rm max}=23$ (see middle left panel of Fig.~\ref{percentage}).

\begin{figure}
\includegraphics[width=7.5cm]{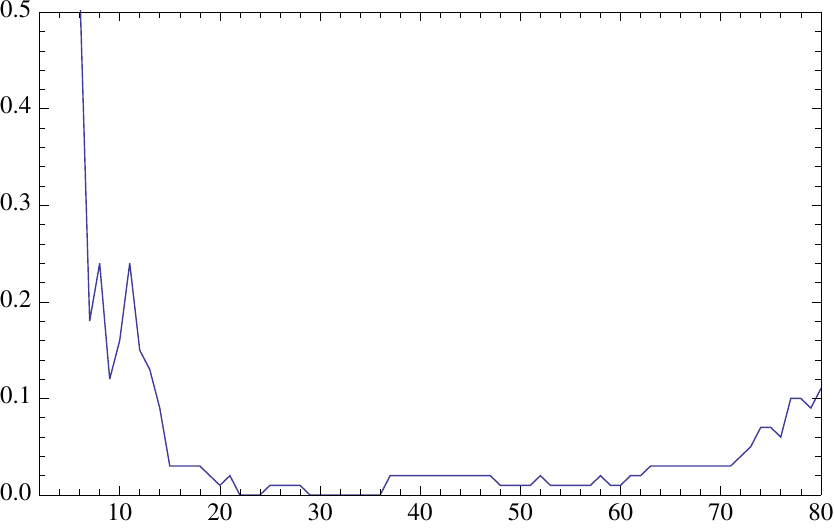}
\includegraphics[width=7.5cm]{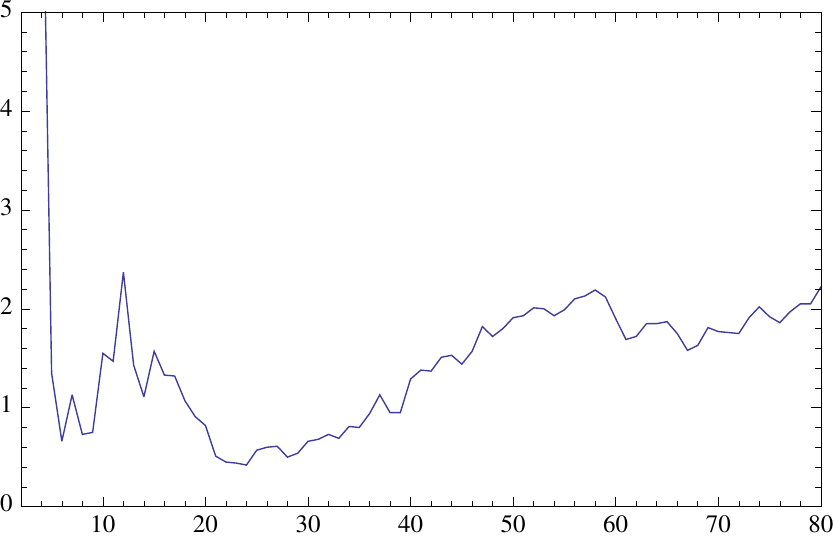} \\
\includegraphics[width=7.5cm]{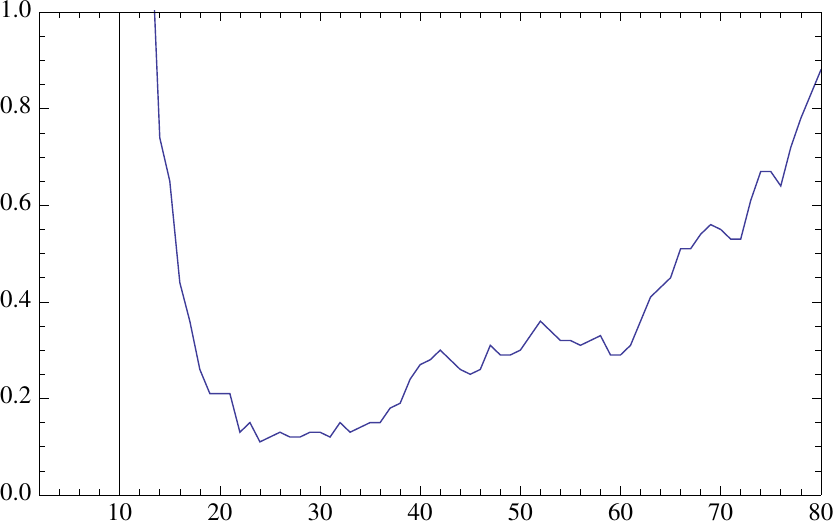}
\includegraphics[width=7.5cm]{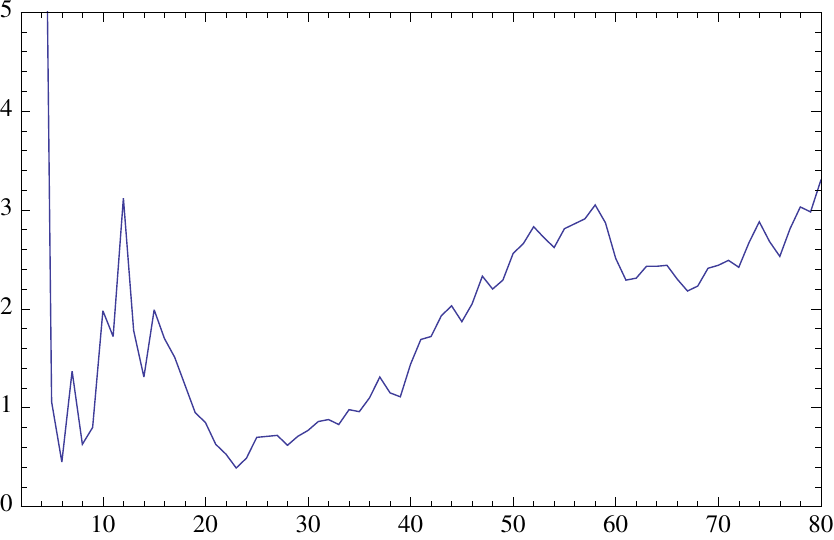} \\
\includegraphics[width=7.5cm]{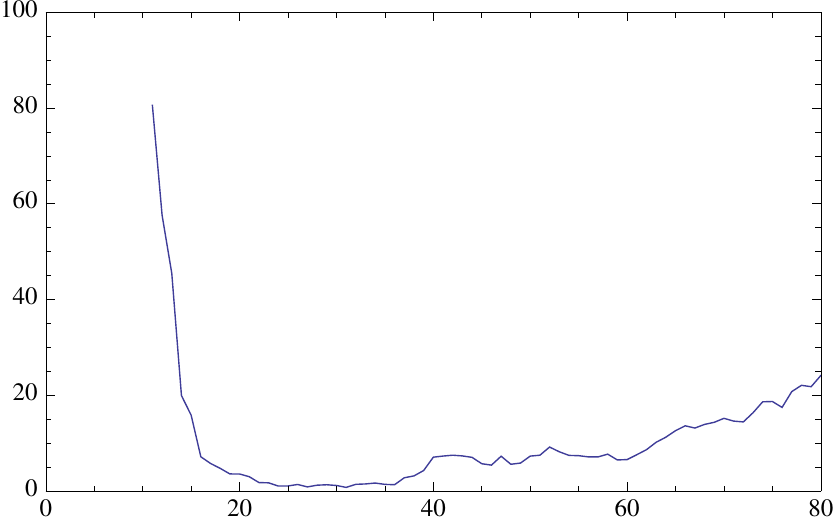}
\includegraphics[width=7.5cm]{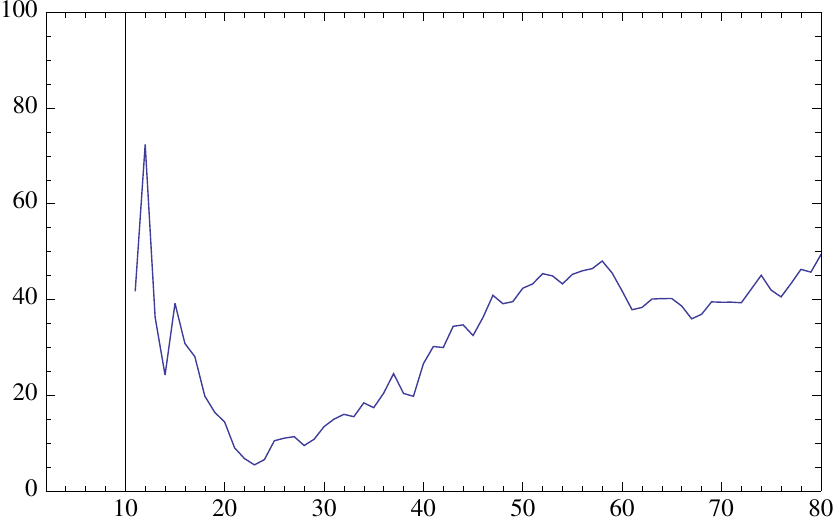} \\
\includegraphics[width=7.5cm]{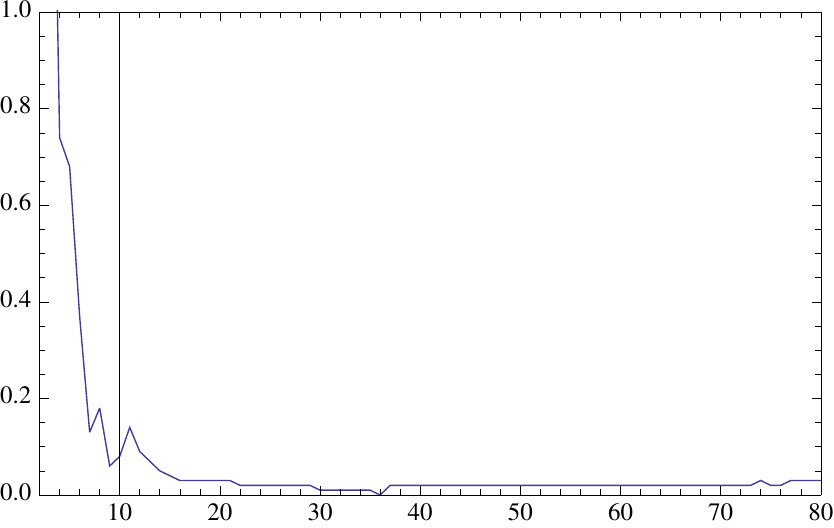}
\includegraphics[width=7.5cm]{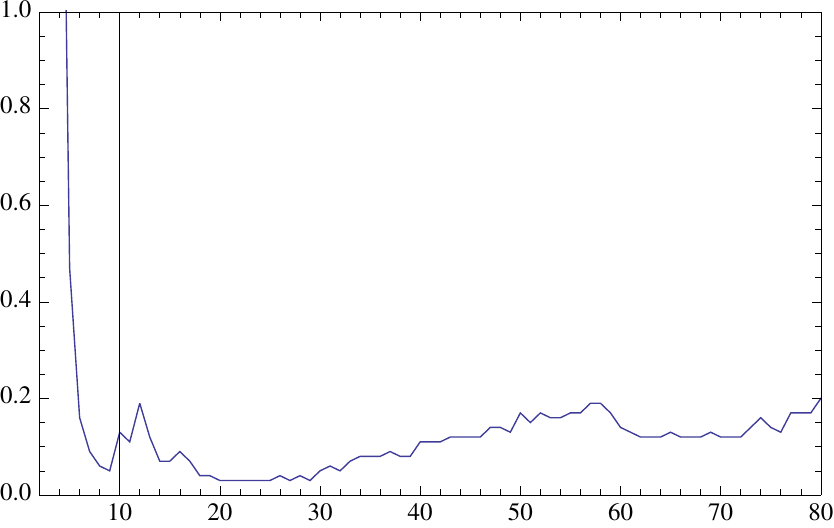}
\caption{
Percentage versus the multipole number $\ell_{\rm max}$ excluding $\ell=2$ (first row), $\ell = 2 \,, 3$ (second row), 
$\ell \le 10$ (third row) and excluding the $m=0$ modes (fourth row).
For each $\ell_{\rm max}$ and for each estimator, 
we plot the percentage to obtain a larger
value w.r.t. Left column is for $S^{+}$ in its anomalous 
direction $\hat{d}_1$, right column is for $S^{-}$ in its anomalous
direction of $\hat{d}_2$.
}
\label{percentage_cut}
\end{figure}

It is important to assess the weight of the low multipoles in quantifying the anomaly of the two directions.
To do so, we have analyzed various cases not including the first multipoles from the analysis. More precisely, we consider three different 
cases, by excluding from the analysis the quadrupole, both quadrupole and octupole and the first 9 multipoles (the sum starts 
from $\ell=11$ in the latter case). The corresponding results are shown in Fig. \ref{percentage_cut}. 
For the $\hat{d}_2$ (symmetry) direction, the quadrupole has an impact in reducing 
the anomaly and by cutting the first 9 multipoles the anomaly almost disappears. Thus, it seems that the quadrupole makes some 
contribution to the effect of anomalous mirror symmetry, but it is not the
dominant one. The reported anomaly here is therefore not just given by the anomalous alignment of the quadrupole and octupole found 
in \cite{Copi:2005ff}. On the other hand, for the $\hat{d}_1$ (anti-symmetry) direction, there is a noticeable change when 
cutting the low multipoles, but an anomalous behaviour still persists.

On summarizing the first three panels of Fig. \ref{percentage_cut}, it is clear that
the anomaly for $S^+$ has a contribution from intermediate multipoles, $10 \lesssim \ell \lesssim 35$,
whereas $S^-$ anomaly is mainly due to low multipoles and almost disappears
for $l>10$.


Since the direction of our $\hat{d}_2$ axis is close to that of the CMB dipole which is mainly kinematic and
local (compared to cosmological scales), the question arises naturally if it may be caused by an 
imperfect dipole subtraction or, more generally, if it is somehow related to the dipole which is, in particular, 
axially symmetric.
Also it is interesting to check if there exists some traces of {\em axial} symmetry with respect to the axes we found
(and which define planes of partial mirror symmetry orthogonal to them). For these purposes, we repeat our
investigations excluding now the axially symmetric component with $m=0$ from the estimators $S^+$ and $S^-$ using
their multipole representation (\ref{defSpemh}) with respect to the $\hat{d}_2$ axis, as well as to the $\hat{d}_1$
axis, too (though in the latter case there is no reason to expect a significant amount of axial symmetry). This
method is similar to that used in the recent paper \cite{NZKC11} in the case of the partial parity symmetry.
Results are presented on Fig. 5, too. It is clearly seen that previous curves for these estimators are practically
unchanged (in fact, the anomaly even becomes slightly more for the $\hat{d}_2$ direction). This clearly shows that
axially symmetric contributions to the anomalies found are small and that the directions $\hat{d}_1$ and $\hat{d}_2$
may not serve as symmetry axes.

\section{Discussion}
\label{discussion}

\begin{figure}[h]
\includegraphics[width=7.8cm]{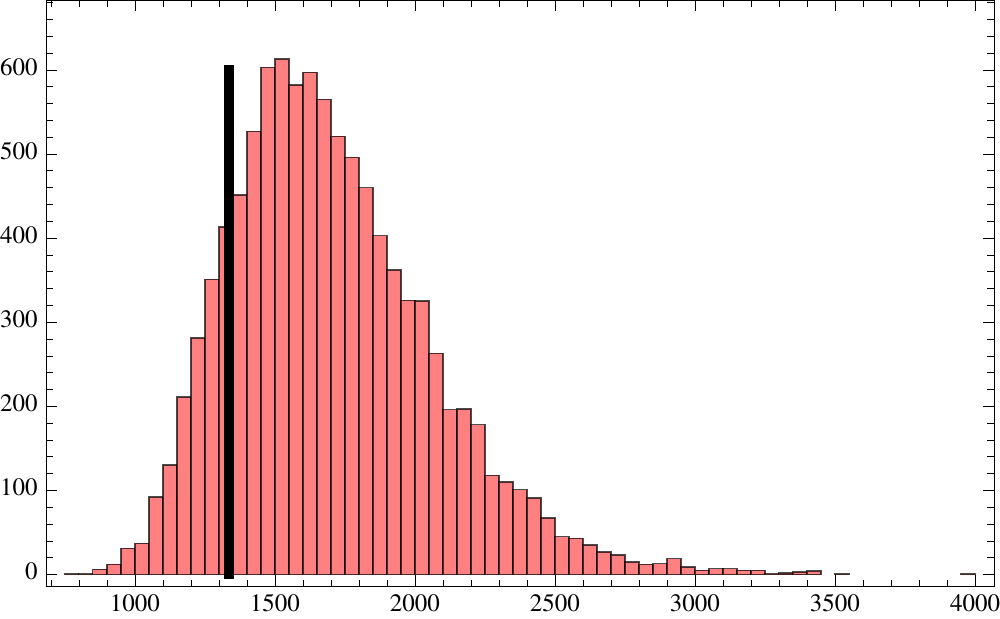}
\caption{Value of the variance of the WMAP 7yr ILC map (thick error bar) compared with its
MC distribution based on 10000 realizations of the WMAP 7 yr $\Lambda$CDM best-fit.}
\label{WMAP7variance}
\end{figure}

One might argue that the weight of the variance term (direction independent) in our estimators $S^\pm$, $Q$ 
in Eqs. (\ref{defQ},\ref{defSpem}) connects our reported anomalies on the mirror-parity directions to the
known problem of the low variance in WMAP maps.
We have computed the low variance of WMAP 7 yr ILC at (with a Gaussian smoothing of $9.1285$ degree)
against MC simulations based on the WMAP 7 yr $\Lambda$CDM best-fit in Fig. \ref{WMAP7variance}:
although we cannot state that the WMAP variance is anomalous at a significant statistical confidence level, it is indeed smaller
that the average value of the MC distribution, as shown in Fig. \ref{WMAP7variance}.

\begin{figure}[h]
\includegraphics[width=4.8cm]{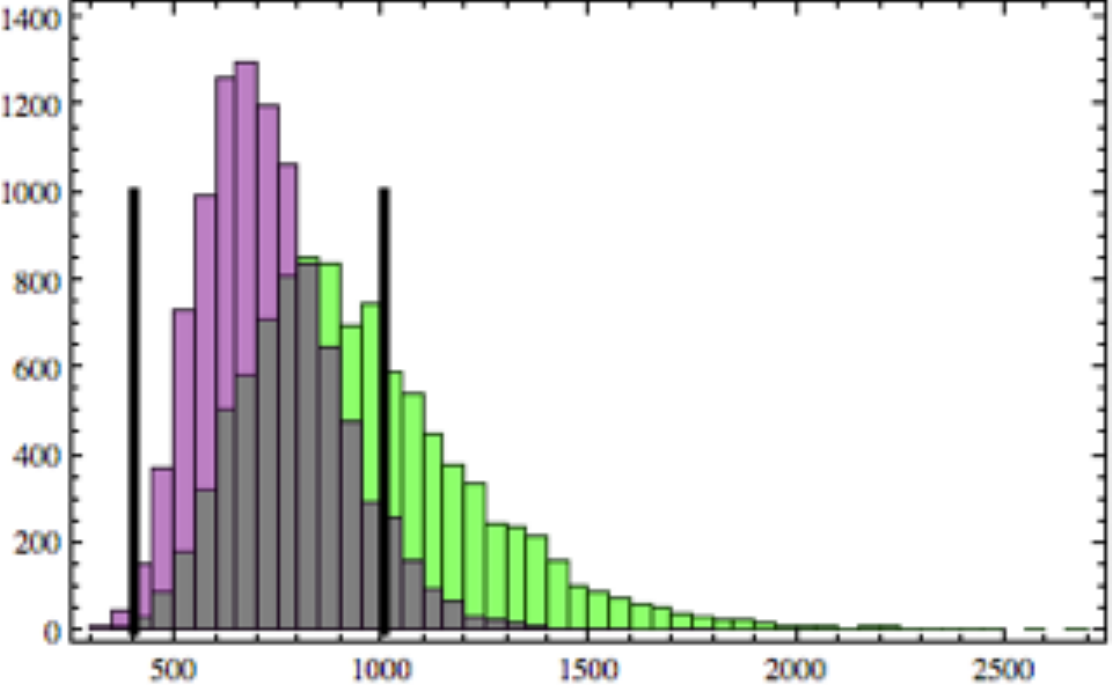}
\includegraphics[width=4.8cm]{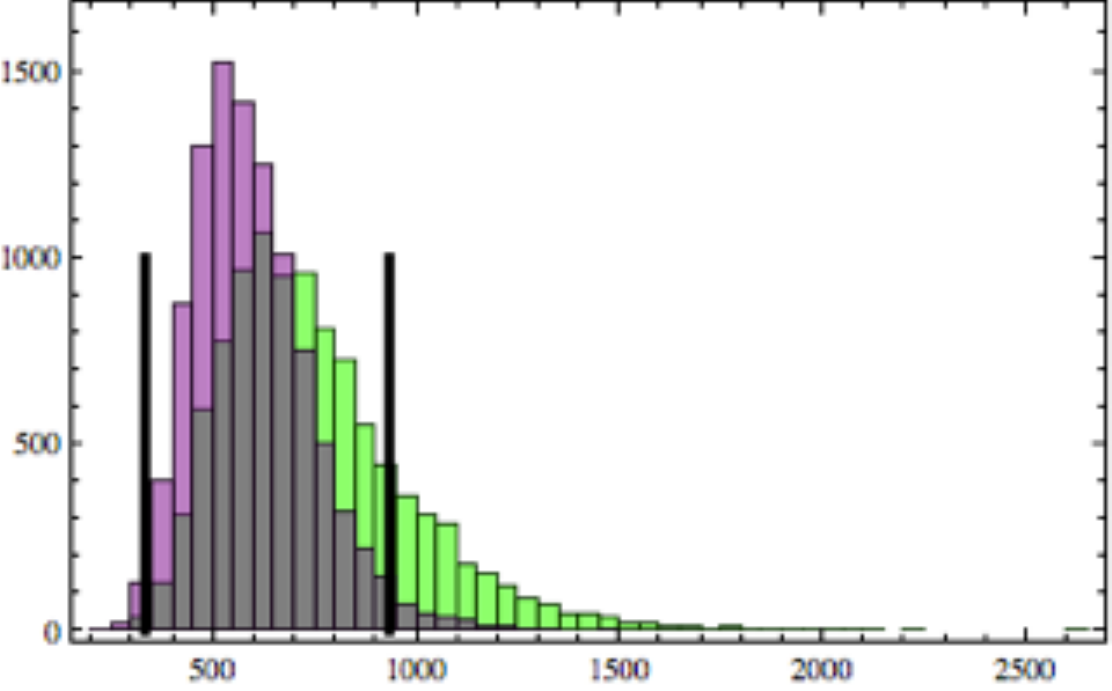}
\includegraphics[width=4.8cm]{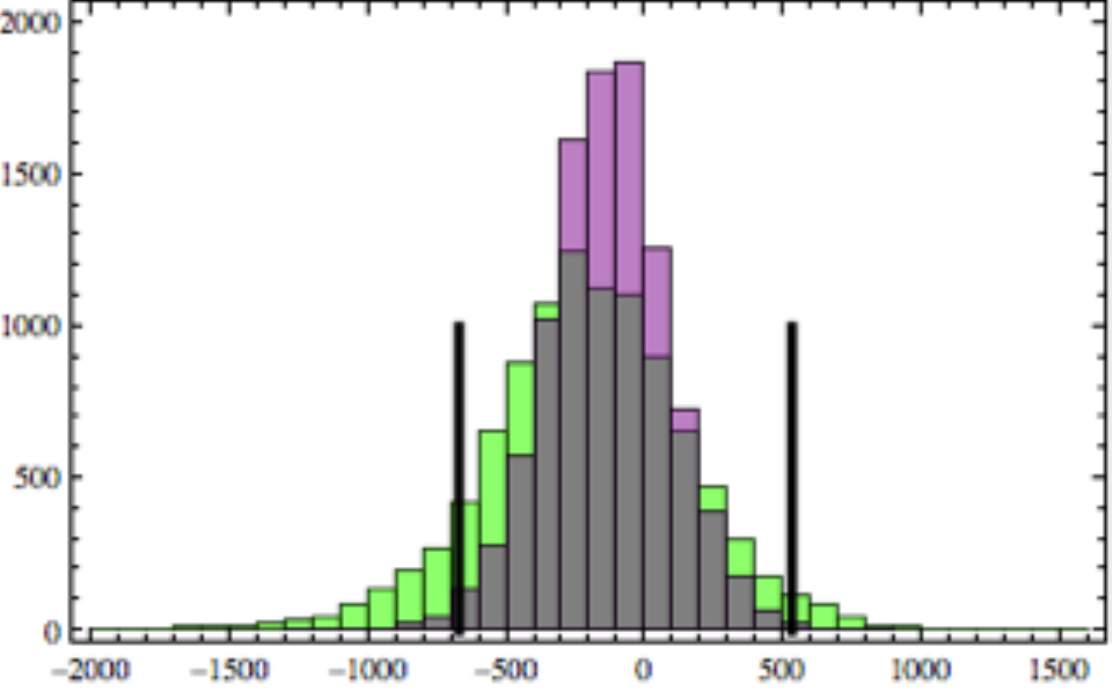}
\caption{Value of the estimators $S^\pm$, $Q$ in the WMAP 7yr ILC map (thick error bars) compared with its
MC distribution based on 10000 realizations of the WMAP 7 yr $\Lambda$CDM best-fit (green histograms)
and of the WMAP 7 yr temperature power spectrum (purple histograms).}
\label{estomatorsvsMCobserved}
\end{figure}

In order to try to disantangle the issue of the low WMAP variance from our findings, we also compare our anomalous values with
10000 MC simulations based on the WMAP 7 yr temperature
power spectrum (publicly available at the NASA Lambda web site) which contains the small values for the quadrupole and octupole,
rather than the theoretical $\Lambda$CDM theoretical best-fit.
The comparison of our results against this second MC shows
that the reported directions for $S^\pm$ are again anomalous, respectively at $99.50 \%$ and $99.17 \%$ confidence level,
but a little less than
with respect to the MC based on the WMAP 7 yr theoretical best-fit, in agreement with
the WMAP low variance shown in Fig. \ref{WMAP7variance}. 
The most intriguing result is that the minimum and maximum values of $Q$ become much more
anomalous with respect to this second MC, respectively at $99.83 \%$ and $99.24 \%$ confidence level:
this fact gives an alternative perspective to our results, now connecting our reported anomalies to the
correlation across the plane, as shown in figure \ref{estomatorsvsMCobserved}.
Table II displays the probabilities with respect to the second MC based on the WMAP 7 yr temperature
power spectrum.

\begin{table}[ht]
\caption{Probabilities (in percentage) to obtain a larger value for the considered directions and pixel based estimators with respect
to the MC with 10000 realizations of the WMAP 7 yr temperature power spectrum} 
\centering 
\begin{tabular}{c c c c c} 
\hline\hline 
Estimator / Direction & $\hat{d}_1$ &  $\hat{d}_2$\\ [0.5ex] 
\hline 
$S^{+}$ & 99.50 & 6.15 \\ 
$S^{-}$  & 2.47 & 99.17 \\
$Q$ & 0.17 & 99.24 \\  [1ex] 
\hline 
\end{tabular}
\label{tableprobabilities2} 
\end{table}

\begin{figure}
\includegraphics[width=6.8cm]{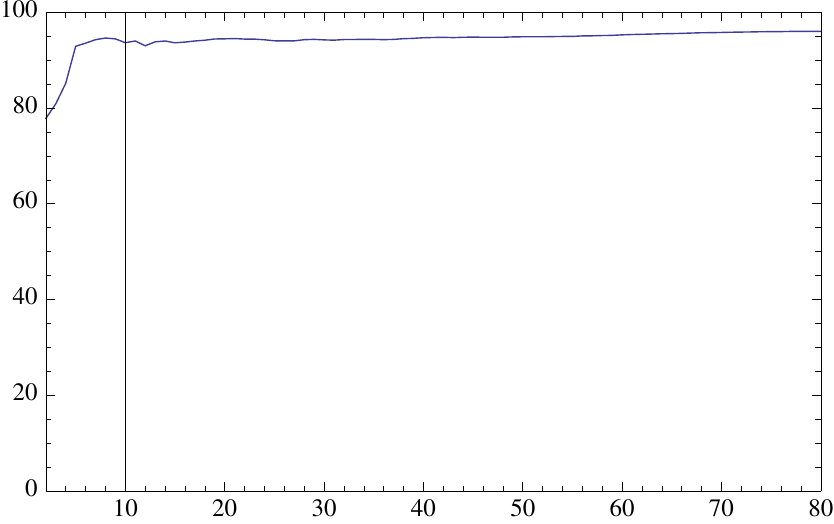}
\includegraphics[width=6.8cm]{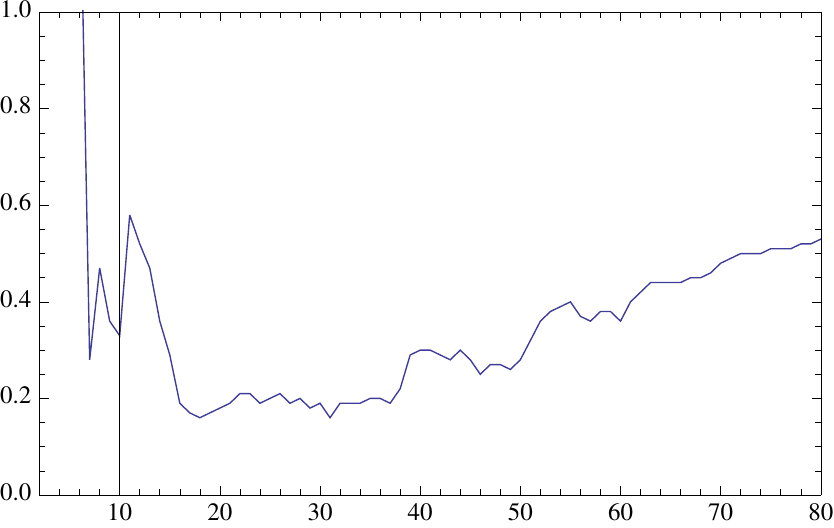} \\
\includegraphics[width=6.8cm]{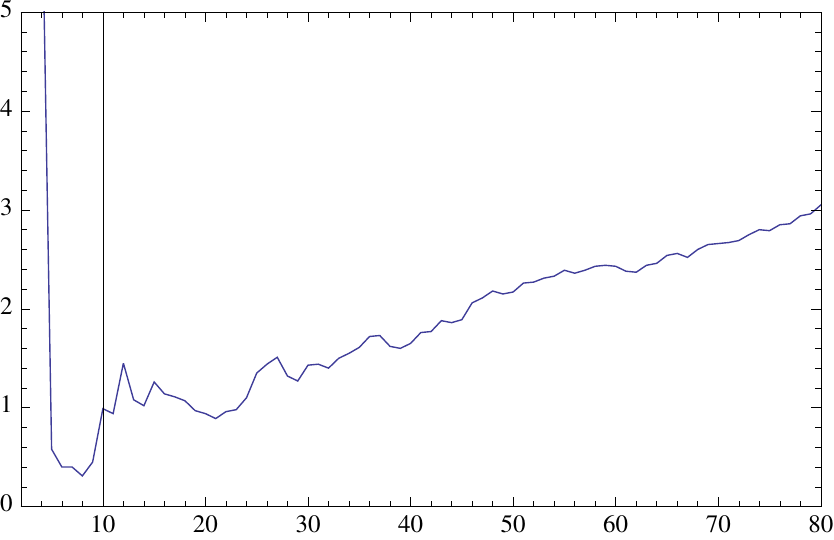}
\includegraphics[width=6.8cm]{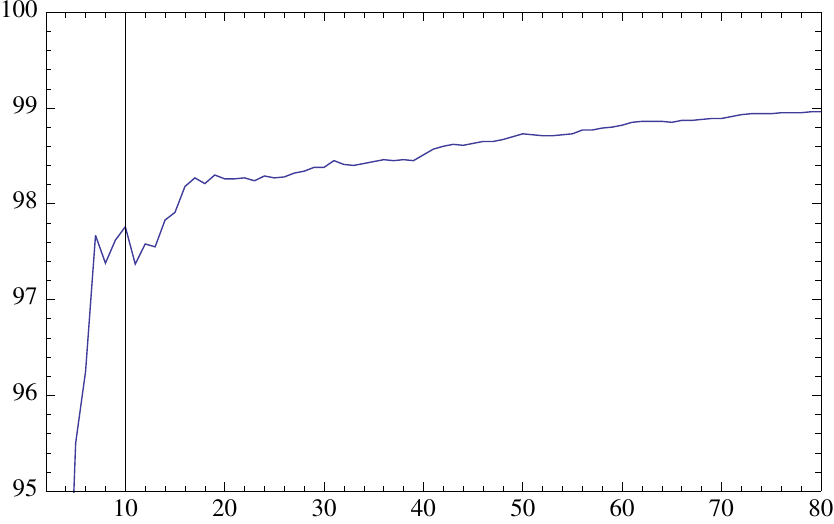} \\
\includegraphics[width=6.8cm]{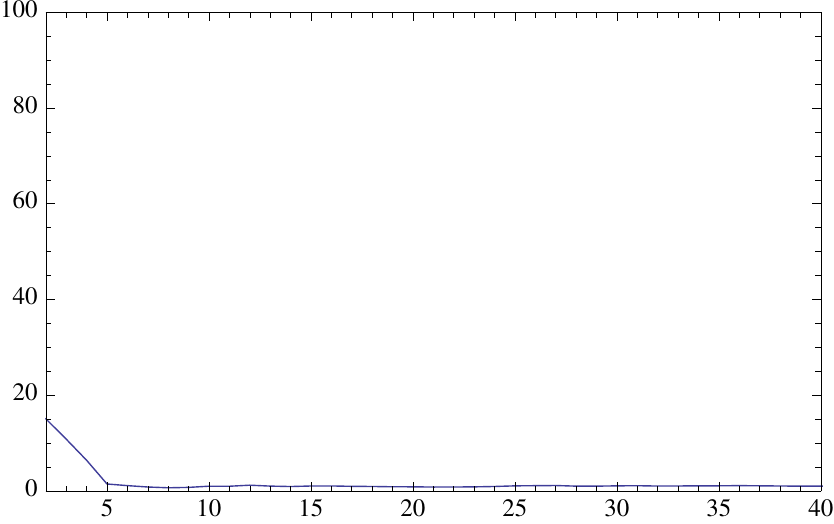}
\includegraphics[width=6.8cm]{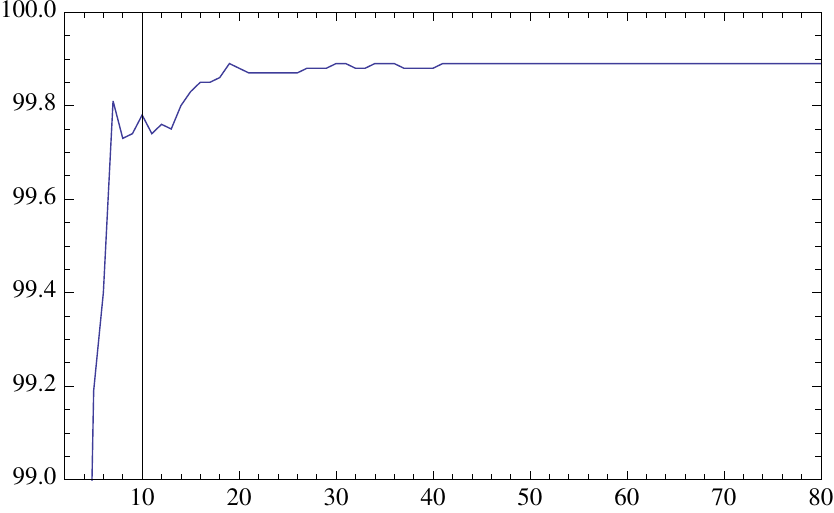}
\caption{Percentage versus the multipole number $\ell_{\rm max}$ with respect to the WMAP 7 ys observed power spectrum.
For each $\ell_{\rm max}$ and for each estimator, we plot the
percentage to obtain a larger value w.r.t. the WMAP ILC one.
Left column is for the $\hat{d}_2~(42^\circ,260^\circ)$ direction and right column is for the $\hat{d}_1~
(107^\circ,264^\circ)$ direction.
First row for $S^{+}$, second row for $S^{-}$ and third row for $Q$.}
\label{percentage_observed}
\end{figure}

We have corroborated our
analysis in pixel space by the analysis in the harmonic domain: we therefore present plots analogous to Fig. 4,
now with respect to the temperature WMAP 7 yr power spectrum, in Fig. \ref{percentage_observed}. 
The plots confirm the analysis in the pixel domain, i.e. that our reported anomalies presented in section 2 are not related 
to the low variance problem of WMAP 7 yr map and that the correlation across $\hat{d}_1$ and $\hat{d}_2$ are anomalous with respect 
to this second MC. This analysis clarifies the anomaly reported with respect to MC simulations which keeps fixed the axis of (anti-) symmetry to the anomalous ones. 
We do not repeat this analysis with respect to simulations which vary the direction in which the maximal (anti-) symmetry is generated.

\section{Conclusions}
\label{conclusions}

We have searched for hidden mirror symmetries in the WMAP 7 year
ILC temperature map. For this aim, we have developed new global
pixel estimators and their relative expressions in the harmonic
domain. We have found that two different axes exist for which the
CMB temperature anisotropy pattern is anomalously symmetric
(anti-symmetric) at the 99.84 (99.96)$\%$ confidence level, if compared to a result for an arbitrary axis 
in simulations without the symmetry. The
angle between them is about $65^{\circ}$. These axes correspond to
the minimum and maximum of CMB temperature correlations with
respect to reflection in a plane. We have verified that the above mentioned 
probability of the anomalies and the position of the axes are
robust to the introduction of the galactic mask. 
The axis
$\hat{d}_1$ for which the amount of mirror anti-symmetry is
maximal is nearly orthogonal to the ecliptic plane, while the axis
$\hat{d}_2$ with the maximal amount of mirror symmetry is close to
the kinematic dipole direction. It is interesting to note that the
$\hat{d}_1$ axis is also close to the one which identifies the N/S
asymmetry \cite{Hoftuft:2009rq}. Removal of CMB temperature
multipoles with $m=0$ with respect to these axes does not change
the anomalous behaviour of our mirror symmetry (anti-symmetry)
estimators showing that the symmetric (anti-symmetric) patterns
found are not axially symmetric with respect to the axes
$\hat{d}_1$ and $\hat{d}_2$. Our findings differ from those
previously published on mirror symmetries
\cite{Gurzadyan07,Gurzadyan:2008ze} using different estimators.
Neither of those axes is close to the preferred direction found in
the recent paper \cite{MHKM11} for a model of anisotropic dark
energy.

If instead the real data are compared to the minimum of $S^\pm$, $Q$ across the sky in each of 10000 random
simulations, the statistical significance drops to $92.39 (76.65) \%$ for $S^+$ ($S^-$) on the WMAP 7 yr ILC map, as can be seen in
Fig. \ref{absolute}; $14.77 \%$ of the simulations
lie to the left of the the minimum value for $Q$ on the WMAP 7 yr ILC. On the basis of this second comparison of absolute minimum values of estimators on
real data and simulations,
the anomaly we find translates into the question of the anomalous alignment between the normals
to the planes of maximal mirror symmetry of real data with known directions such as the normal to the ecliptic plane and the cosmological dipole.

To clarify the nature of these anomalies, we have investigated the weight of the low multipoles in these new
anomalies by exploiting the estimators in multipole space: 
by removing the quadrupole and octupole from the analysis, we have shown 
that the (anti-) mirror symmetries reported here are novel results and are not simply
a different perspective of the previously discovered anomalies in the quadrupole and octupole in intensity power 
spectrum and multipole vectors \cite{Copi:2005ff}. 
By cutting $\ell \le 10$, the statistical significance
of the anomaly in both directions drastically drops. However, whereas the anomaly in $S^+$ (its anomalously low
value of for the $\hat{d}_1$ direction) remains though becomes less pronounced, the anomaly in $S^-$ for the
$\hat{d}_2$ direction almost disappears. In our opinion, this different behaviour might be not completely
surprising: the direction $\hat{d}_1$ is close to that of the N/S asymmetry which has been shown to be present
also at intermediate multipoles of the order of $\ell \sim 100$ \cite{Hoftuft:2009rq}. This suggests that the
(anti-symmetric with respect the mirror reflection) anomaly in $S^+$ and the N/S asymmetry may have the same and
probably secondary (even local) origin, that does not prevent it to be a real physical, non-systematic effects. 
On the opposite, the symmetry anomaly which we find in $S^-$ is due only
to low multipoles and therefore it may have a cosmological or even primordial origin. 
Its anomalous alignment with the cosmic dipole may be a sign of their common origin, too.
Whatever the origin
of these anomalies is, these two new anomalies differ in this way.

We have shown that these hidden mirror symmetry properties have little to do with the low variance in 
WMAP maps. To gauge away the issue of the low WMAP variance from our results, we have also compared our findings with 
10000 MC simulations based on the WMAP 7 yr temperature
power spectrum, which contain the low quadrupole value. With respect to this second MC,  
the anomalies for $S^\pm$ do not disappear, whereas the anomaly for $Q$ becomes more evident and connects our 
reported anomalies to the correlation across the plane.

While the present paper was being prepared for submission, the article \cite{BenDavid:2011fc} appeared where a
search for the mirror (anti-)symmetry in CMB fluctuations was also performed. In contrast to our paper, the analysis
in \cite{BenDavid:2011fc} is restricted only to low multipoles up to $\ell_{\rm max}=7$ for an estimator defined in the
harmonic domain which is similar to $Q$, but with a $\ell$ dependent weight. The statistically significant anomaly
reported in \cite{BenDavid:2011fc} is an anti-symmetry anomaly in the direction $(\theta=109^\circ,\phi=266^\circ)$
which is close to our direction $\hat{d}_1$. We stress that we do not report any anomaly for $Q$ in the WMAP 7 yr
ILC map, but the different estimator and different data analysis adopted here and in \cite{BenDavid:2011fc} do not
allow a direct comparison of the results.

So, summarizing, we have found two planes, with respect to
reflection in which CMB angular temperature fluctuations possess
the most pronounced and anomalous amount of partial mirror
symmetry or anti-symmetry. From these two cases, the mirror
symmetry one for which the corresponding plane has the normal
direction $\hat{d}_2$ is more promising for the purpose of
extracting new cosmological information about the present and
early Universe since it is almost entirely due to low multipoles
$\ell$. It is clear that the amount of existing data is not
sufficient to make more definite conclusions about the origin of
this anomaly. The forthcoming release of WMAP 9 year data and then
the independent observations from {\it Planck} will shed more
light on these new mirror anomalies. Additional independent
observations or different data analysis will be cross-checks of
the importance of possible systematics, as beam asymmetries,
foreground residuals and subtleties related to the kinematic
dipole in the mirror symmetries found in this paper.

\begin{acknowledgments}
We thank Glenn Starkman for useful comments and the anonymous referee for 
critical remarks, anwers to which have contributed to improve our manuscript.
We acknowledge the use of the Legacy Archive for Microwave
Background Data Analysis (LAMBDA). Support for LAMBDA is provided
by the NASA Office of Space Science. Some of the results in this
paper have been derived using the HEALPix \cite{gorski} package.
This work has been done in the framework of the {\it Planck} LFI
activities. This work has been partially supported by the ASI
contracts Planck LFI Activity and I/016/07/0 COFIS. AS was
partially supported by the Russian Foundation for Basic Research
(RFBR) under the grant 11-02-12232-ofi-m-2011 and by the Research
Programme "Astronomy" of the Russian Academy of Sciences.
\end{acknowledgments}

\appendix

\section{Multipole expansion of the pixel based estimators}
\label{appendix}

In this Appendix we give the relevant algebra to Eqs. (2.4-8). We consider the expansion of temperature anisotropies
in spherical harmonics
\be
\frac{\delta T}{T} (\theta , \varphi) = \sum_{\ell \, m} a_{\ell m} Y_{\ell m} (\theta, \varphi)
\ee
with the following convention
\be
Y_{\ell m} (\theta, \varphi) = \left( \frac{2 \ell + 1}{4 \pi} \frac{(l-m)!}{(l+m)!} \right)^{1/2}
P_\ell^m (\cos \theta) e^{i m \varphi} \,.
\label{Ydef}
\ee
The convention on the averages is
$N_{\rm pix}^{-1} \sum_{j=1}^{N_{\rm pix}} = \frac{1}{4 \pi} \int d \Omega$.
By using the multipole expansion, Eq. (2.2) and the completion property of the spherical harmonics
in Eq. (2.2), we obtain:
\be
S^{\mp} (\hat n_i) &=& \frac{1}{2 N_{\rm pix}} \sum_{j=1}^{N_{\rm pix}}
\left( {\delta T \over T} (\hat n_j) \right)^2 + \left( {\delta T \over T} (\hat n_k) \right)^2
\mp {\delta T \over T} (\hat n_j) {\delta T \over T}(\hat n_k) \\
&=& \frac{1}{8 \pi} \left[ \sum_\ell (2 \ell +1) C_\ell
\mp \int d \Omega {\delta T \over T} (\theta,\varphi) {\delta T \over T}(\theta, \pi -\varphi) \right]
\label{algebraSpem}
\ee
where we have used:
\be
C_\ell = \frac{1}{2 \ell +1} \sum_{m=-\ell}^\ell |a_{\ell m}|^2 \,.
\ee
The last term in Eq. (\ref{algebraSpem}) is simply $-Q$ and can be rewritten as:
\be
\int d \Omega {\delta T \over T} (\theta,\varphi) {\delta T \over T}(\pi - \theta, \varphi)
&=& \sum_{\ell m} \sum_{\ell' m'} a_{\ell m} a^*_{\ell' m'} (-1)^{\ell' + m'}
Y_{\ell m} (\theta,\varphi) Y_{\ell' m'}^* (\theta,\varphi) \\
&=& \sum_{\ell m} (-1)^{\ell+ m} |a_{\ell m}|^2 \\
&=& \sum_{\ell} (\ell +1) C_\ell^{+} - \ell C_\ell^{-} \,,
\label{algebraQ}
\ee
where $C_\ell^+ \,, C_\ell^-$ are defined below Eq. (2.5).

\end{document}